\documentstyle[12pt,psfig,amssymb]{article}
\begin{document}
\begin{center}

{\Large Soliton-radiation coupling in the parametrically driven, damped
nonlinear Schr\"odinger equation}\\[.5cm]


{V. S. Shchesnovich\footnote{valery@maths.uct.ac.za}
and I. V. Barashenkov\footnote{igor@cenerentola.mth.uct.ac.za}\\[.5cm]
{\it Department of Mathematics and Applied Mathematics, University of \\
Cape Town, Private Bag Rondebosch 7701, South Africa.}\\[.5cm]
}
\end{center}
\newpage

\begin{abstract}

We use the Riemann-Hilbert problem to study the interaction of the 
soliton with  radiation in the parametrically driven, damped nonlinear 
Schr\"odinger  equation. The analysis is reduced to the study of
a finite-dimensional dynamical system for  the amplitude and phase
of the soliton and the complex amplitude of the long-wavelength radiation.
In contrast to previously utilised Inverse Scattering-based perturbation 
techniques, our approach is valid for arbitrarily large driving strengths and
damping coefficients. We show that, contrary to suggestions made
in literature, the  complexity observed in the soliton's dynamics
cannot be accounted for just by its coupling  to the long-wavelength
radiation.

\end{abstract}
\newpage

\section{Introduction}
\label{first}
A variety of
nonlinear wave phenomena in one dimension  can be modelled
by the perturbed nonlinear Schr\"{o}dinger (NLS) equation:
\begin{equation}
iq_t+q_{xx}+2|q|^2q= {\cal R}(t,x,q,q^*,q_x,q^*_x,...).
\label{1}
\end{equation}
(Here and below the asterisk denotes complex conjugation.)
The present paper  deals with the parametrically driven, damped
NLS, for which 
\begin{equation}
{\cal R} =he^{2i \Omega_0  t}q^*-i\gamma q;
 \quad h, \;  \Omega_0 ,\;\gamma>0.
\label{pert}
\end{equation}
This equation describes
nonlinear Faraday resonance in a vertically oscillating water
trough~\cite{F1,Meron,FA3,F8,F3,F7,F4,F5,F6,Il'ichev,FA1,astruc};
an easy-plane ferromagnet with a combination of a stationary and a
high-frequency magnetic field in the easy plane \cite{M5};
and the effect of  phase-sensitive  amplifiers on
solitons propagating in optical
fibres~\cite{Op1,Op2,Op3}.

The equation (\ref{1})-(\ref{pert}) has two stationary soliton solutions,
one of which is unstable for all $h$ and $\gamma$ and hence  usually
disregarded \cite{M5}. (The frequency $\Omega_0$ can always be
scaled to unity leaving  $h$ and $\gamma$ as the only two control
parameters.)  The other stationary soliton is stable for small values
of $h$ but looses its stability to a periodically oscillating soliton
as $h$ is increased above a certain critical value
$h=h_c(\gamma)$, for the fixed $\gamma$ \cite{M5}.
The chart of attractors arising as $h$ is  increased further, was compiled 
in \cite{Bondila} and can be summarised as follows. 
If $\gamma$ is small, the oscillating soliton undergoes a 
sequence of bifurcations which culminate in a soliton whose amplitude, width 
and phase are changing chaotically in time. 
For an even greater $h$ (and the same, fixed, $\gamma$)  
the soliton breaks up and decays to zero whereas increasing the driver's 
strength still further, a spatio-temporal chaotic state sets in. On the 
other hand, if $\gamma$ is large, the bifurcations of the soliton's period 
and its breakup are not observed; instead, as $h$ is increased for this
$\gamma$, the periodically oscillating soliton yields
directly to the spatio-temporal chaos. 

Some insight into the mechanism controlling the soliton's transformations 
was gained in Ref.\cite{add2} where a reduced amplitude equation was derived 
for the perturbation of the stationary soliton. Its analysis demonstrated 
that the emission of  radiation waves plays a major role in the soliton's 
dynamics. In particular, the soliton-radiation interaction accounts for the 
stable periodic oscillations of the damped soliton ($\gamma \neq 0$) and for 
the absence of the periodicity in the undamped situation, $\gamma =0$. 
However, the  analysis of \cite{add2} was confined to a neighbourhood of the 
instability threshold $h=h_c(\gamma)$  and hence the reduced amplitude 
equation could not reproduce the entire complexity  in the soliton's 
dynamics which is observed in  numerical simulations with larger $h$ 
\cite{Bondila}.

In the present work we  study  the
soliton-radiation interaction from a different perspective.
Our decomposition of the phase space into soliton and radiation modes will
be based not on the linearisation about the stationary soliton
in the ``$q$-space" (which was the approach of Ref.~\cite{add2}),  but on 
the analysis of the spectral data of the Riemann-Hilbert problem associated 
with the unperturbed NLS equation. This will allow us to examine the role 
of the soliton-radiation coupling for {\it arbitrary\/} $h$ and $\gamma$, 
and not just in the neighbourhood of the instability onset.

There is a number of perturbation schemes   available in
literature which exploit the proximity of the perturbed NLS 
(\ref{1}) to the completely integrable case of the ``pure" NLS, 
eq.(\ref{1})
 with ${\cal R}=0$. (See \cite{7,8,9,11,14,16,Prokopenya,18}  and 
a review \cite{12}). 
Solutions of equation (\ref{1}) have their images in the space of spectral 
data for {\it any\/} ${\cal R}$, of course, but this is of little use in the 
general case. The difficulty here is that
 the evolution equations for the spectral data 
involve the associated eigenfunctions (or, 
equivalently, solutions of the Riemann-Hilbert problem).
In order to obtain a {\it closed\/} 
system of equations for the spectral data, one assumes 
 that  ${\cal R}$ is 
small in some sense and expands the spectral data and  eigenfunctions 
in powers of the small parameter \cite{7,8}.  In the {\it adiabatic\/} 
approximation, for example,
one first derives equations of the (slow) evolution of the 
soliton's parameters (ignoring the radiation completely) and  then 
calculates the spectral density of the radiation emitted by this soliton
\cite{8,9,12}. 
The back-reaction of the radiation on the soliton is not taken into account 
in this approach, therefore. The adiabatic approximation  is 
capable of capturing some 
basic  essentials of the soliton's dynamics, such as the phase-locking of 
the soliton to the periodic driver (see e.g.~\cite{9})
or stability against  perturbations of 
its parameters \cite{8}, and is usually sufficient for {\it very\/} small right-hand 
sides in (\ref{1}).
(See \cite{12} for details.)
 However, it becomes inadequate for somewhat larger 
${\cal R}$,  where the  back-reaction of the emitted radiation on the 
soliton cannot be disregarded \cite{14,16,add1}. 
(For example, the adiabatic approximation   
does not capture the oscillatory instability of the parametrically driven 
soliton  which sets in for $\gamma=0$ and $h$ as 
small as 0.064 \cite{M5}.)

An attempt to go beyond the adiabatic approximation and 
consider the radiation degrees of freedom on equal footing with those of 
the soliton was made by the authors of Ref.~\cite{add1}
whose finite-dimensional reduction included the complex
amplitude of the $k=0$-radiation.
However, their derivation was not entirely 
self-consistent; in particular, their approach produced
an equation for the radiation part of the spectral data which did not 
contain a damping term.  To  reconcile conclusions of their 
finite-dimensional analysis with direct numerical  simulations of the full 
partial differential equation, the  authors had to add the damping in an 
{\it ad hoc\/} way. The amplitude of the radiation wave also remained 
undefined and had to be chosen so as to match the numerics \cite{add1}. 

Similarly to Ref.~\cite{add1}, the purpose of the present work is to study 
the effect of the soliton-radiation coupling on the internal dynamics of the 
damped-driven soliton.  
However, unlike the analysis of Ref.~\cite{add1} and 
perturbation schemes appeared 
elsewhere, our approach is {\it not\/} using the assumption of the smallness 
of ${\cal R}$. Instead, we will exploit the fact 
that the stationary soliton of equation (\ref{1})-(\ref{pert}) with ${\cal 
R} \neq 0$ coincides --- up to a simple phase transformation --- with the 
soliton solution of equation (\ref{1}) with ${\cal R}=0$. Consequently, the
stationary soliton of the parametrically driven damped NLS
with arbitrarily large $h$ and $\gamma$ can be associated with a stationary 
zero of the Riemann-Hilbert problem underlying the ``pure", integrable,  
NLS equation. This observation allows to choose a different small 
parameter; instead of the smallness of $h$ and $\gamma$ we will be utilising 
the proximity of the solution in question to the stationary soliton {\it of 
the perturbed equation\/}. (Another
assumption that we  are going to make, following  \cite{add1},
is that the radiation is {\it linear\/}, i.e. 
it couples to the nonlinearly evolving soliton but does not 
interact with itself.) 
 Using this small parameter we will be able to 
obtain a closed  system of equations describing, approximately, the 
evolution of the spectral data.

Some analytical insights can be gained already from the linearisation of 
this system (about the zero of the 
Riemann-Hilbert problem corresponding to a single soliton). 
We will show that the linearisation explains the origin of the oscillatory 
instability 
which serves as a starting point of the sequence of
secondary bifurcations and leads to the emergence of the increasing 
complexity in the soliton's dynamics.   (So far, the oscillatory instability 
and related Hopf bifurcation    remained  just facts of numerical analysis 
\cite{M5},\cite{add2}.) It also indicates that  
the soliton interacts most intensively with radiation waves near  the lower 
boundary of their spectrum  (where $k=0$). This seems to be in agreement 
with earlier suggestions --- made for a closely related {\it externally\/} 
driven NLS ---  that keeping just the $k=0$ mode is sufficient to capture 
the basic features of the partial differential equation 
\cite{add1,TSFGR,Birnir}.  Accordingly,  we  focus
 our subsequent
efforts on the verification of this hypothesis. Namely, we explore the 
effect of the coupling to the $k=0$ radiation on the {\it nonlinear\/} 
dynamics of the soliton.  Keeping only infinitely long waves allows to 
obtain  a four-dimensional system for the soliton phase  and the complex 
amplitude of the radiation. Results of the analysis of this system are then 
 compared to the phenomenology of the parametrically driven
soliton reported in literature. 
 We will show that taking 
the $k=0$  radiation into account can explain  {\it some\/} 
dynamical effects, most notably the occurrence  of  the soliton's 
breakup and decay to zero. We will also identify  aspects of the 
behaviour  that {\it cannot\/} be attributed just to the coupling of the
soliton to the
long-wavelength radiation
 and therefore require  invoking other degrees 
of freedom.  These aspects will include, in particular,  the shape of the  
instability domain on the $(h,\gamma)$-plane and the route to the
(temporal) chaos.

The paper is organised as follows. The next section contains a brief  
summary of the Riemann-Hilbert problem and the perturbation theory based
upon 
it. Section  \ref{adiabat} discusses the adiabatic approximation and its 
shortcomings. In section \ref{second} we linearise 
the evolution equations for
the spectral data, while the 
{\it nonlinear\/} four-dimensional system for the soliton's parameters and the 
complex amplitude of the $k=0$ radiation  is derived in section \ref{third}. 
Numerical simulations of this system are reported in section \ref{fourth}. 
The last section (section \ref{fifth}) contains conclusions of  our analysis.

\section{Inverse Spectral Transfrom for the ``pure" and perturbed NLS 
equation}
\label{maths}

In this section we review the main points of the modern version of the 
Inverse Scattering Transform, known as the 
method of the Riemann-Hilbert problem, for the NLS equation. After that we 
outline the basic principles of the Inverse Scattering-based perturbation 
theory, in its particular Riemann-Hilbert formulation. 

\subsection{The Riemann-Hilbert problem}
\label{rh}
The applicability of the Inverse Scattering Transform to the unperturbed NLS 
equation (eq. (\ref{1}) with  ${\cal R} =0$) is due to the fact that the NLS 
serves as the compatibility condition for the following system of two linear 
equations for the matrix-valued function  $\Psi(x,t;\zeta)$:
\label{2}
\begin{equation}
\partial_x\Psi= i\zeta[\sigma_3, \Psi]+iQ\Psi,
\label{2a}\end{equation}
\begin{equation}
\partial_t \Psi=-2i\zeta^2[\sigma_3, \Psi]-\left(2i\zeta Q+\sigma_3 
Q_x-iQ^2\sigma_3\right)
\Psi,
\label{2b}\end{equation}
where the potential matrix 
\[
Q =\left(\begin{array}{cc}0 & q \\ 
q^* & 0\end{array}\right);
\]
$q=q(x,t)$ is a solution of the unperturbed NLS and $\sigma_3$ is the Pauli 
matrix \cite{4}. Knowing $\Psi$, the potential $Q$ can be recovered from the 
asymptotic expansion of $\Psi(\zeta)$ as $\zeta\to\infty$:
\begin{equation}
\Psi({\zeta})={I}+\Psi^{(1)}{\zeta}^{-1}+\ldots,\quad 
Q=[\Psi^{(1)},\sigma_3],
\label{qlim}\end{equation}
whereas $\Psi$ is found via the analytic factorisation in the complex 
$\zeta$-plane. The factorisation problem is known as the matrix 
Riemann-Hilbert problem; the interested reader may consult 
Refs.~\cite{5,6,new1,new2} for the full account of this technique while here 
we only give a brief summary.

First one defines the  Jost solutions of equation (\ref{2a}) by their 
asymptotic behaviour as \mbox{$x\to\pm\infty$:} $J_\pm\to I$. Then the 
matrix-valued  function 
\begin{equation}
\Psi_+({\zeta})=\left(\!\begin{array}{cc}(J_+)_{\cdot 1}({\zeta})
&(J_-)_{\cdot 2}({\zeta})\end{array}\!\right)
\label{4}\end{equation}
is a solution to (\ref{2a})-(\ref{2b}) holomorphic in the upper half 
of the complex ${\zeta}$-plane (Im${\zeta}\ge0$). In (\ref{4}), 
$(J_\pm)_{\cdot l}$ stands for the $l$-th column of  $J_\pm$. 
Noting that the linear problem (\ref{2a})-(\ref{2b}) admits an involution 
$\Psi^\dagger({\zeta})=\Psi^{-1}(\zeta^*)$,  we introduce a matrix function 
holomorphic in the lower half-plane:
\begin{equation} \Psi_-^{-1}({\zeta})=\Psi_+^\dagger(\zeta^*)=
\left(\!\begin{array}{cc}(J_+^{-1})_{1\cdot}({\zeta}),
&(J_-^{-1})_{2\cdot}({\zeta})\end{array}\!\right)^T,
\label{5}\end{equation}
where $({J_\pm}^{-1})_{l\cdot}$ denotes the $l$-th row of the matrix
${J_\pm}^{-1}$ and  $T$ indicates transposition. 
The functions $\Psi_+(\zeta)$ and $\Psi_-(\zeta)$, solutions to equation 
(\ref{2a}), can be expressed through the Jost solutions and the 
elements of the scattering matrix, defined as
\begin{equation}
S=e^{-i{\zeta}x\sigma_3}J^{-1}_+J_-e^{i{\zeta}x\sigma_3}.
\label{6}\end{equation}
Introducing the upper and lower-triangular matrices $S_\pm$,
satisfying the equation $S_+=SS_-$, by
\begin{equation}
S_+=\left(\begin{array}{cc}
1&S_{12}\\
0&S_{22}\end{array}\right),\quad 
S_-=\left(\begin{array}{cc}(S^{-1})_{11}&0\\
(S^{-1})_{21}&1\end{array}\right),
\label{7}\end{equation}
we have 
\begin{equation}
\Psi_+=J_+e^{i{\zeta}x\sigma_3}S_+e^{-i{\zeta}x\sigma_3},
\quad \Psi^{-1}_-=e^{i{\zeta}x\sigma_3}S_+^\dagger 
e^{-i{\zeta}x\sigma_3}J_+^{-1}.
\label{8}\end{equation}

This leads to  the Riemann-Hilbert problem of finding the matrix-valued 
functions $\Psi_+(\zeta)$ and $\Psi_-^{-1}(\zeta)$, holomorphic in the upper 
and lower half-plane of $\zeta$, respectively, and satisfying 
\begin{equation}
\Psi_-^{-1}\,\Psi_+= e^{i{\zeta}x\sigma_3}Ge^{-i{\zeta}x\sigma_3},
\label{9}\end{equation} 
on the real line \cite{6}. Here $\Psi_\pm\rightarrow{I}$ as 
${\zeta}\rightarrow\infty$ and \[
G=\left(\begin{array}{cc}1&g\\
g^*&1\end{array}\right),
\]
with $g(\zeta)=S_{12}({\zeta})$. 
The $t$-dependence of $g$ follows from  
equations  (\ref{9}) and (\ref{2b}). We have $G_t=-2i{\zeta}^2[\sigma_3,G]$, 
hence $g_t = -4i\zeta^2 g$. 
 
If the $\det\Psi_+({\zeta})$ has zeros in its analyticity domain, 
the Riemann-Hilbert problem is said to be singular (or with zeros). Owing to 
the involution, zeros of $\det\Psi^{-1}_-({\zeta})$ are  complex conjugates 
of those of  $\det\Psi_+({\zeta})$. The solution to 
the Riemann-Hilbert problem with zeros can be written as
\begin{equation}
\Psi_\pm({\zeta})={\Psi_0}_\pm({\zeta})\Gamma({\zeta}),
\label{10}\end{equation}
where $\det{\Psi_0}_\pm({\zeta})\ne0$ and the rational matrix function 
$\Gamma({\zeta})$  (the  ``dressing matrix") has zeros of 
$\det\Psi_+(\zeta)$ and poles at the zeros of $\det\Psi^{-1}_-(\zeta)$.  
In the case of $N$ simple zeros ${\zeta}_j$, $j=1,\ldots,N$ (i.e.  
$\det\Psi_+({\zeta})={\cal O}({\zeta}-{\zeta}_j)$ as ${\zeta}\rightarrow 
{\zeta}_j$), the  dressing matrix has the following
structure (see also \cite{6}): 
\begin{equation} 
\Gamma({\zeta})={I}-\sum_{j,l=1}^N\frac{\mid{p_j}\rangle(D^{-1})_{jl} 
\langle{p_l}\mid}{{\zeta}-\zeta^*_l}, \label{11}\end{equation}
where $D_{ln}=({\zeta}_n-\zeta^*_l)^{-1}\langle{p_l}\mid 
p_n\rangle$ and $\langle{p_l}\mid{p_n}\rangle=(p^*_l)_1(p_n)_1
+(p^*_l)_2(p_n)_2$. The vector-columns $\mid{p_j}\rangle$
and vector-rows $\langle{p_j}\mid$ are defined by 
\begin{equation}
\Psi_+({\zeta}_{j})\mid{p_{j}}\rangle=0,\quad \langle{p_{j}}\mid
\Psi_-^{-1}(\zeta^*_{j})=0.
\label{12}\end{equation}
The involution property  gives  
$\langle{p_{j}}\mid=\mid{p_{j}}\rangle^\dagger$ and 
$\Gamma^{-1}({\zeta})=\Gamma^\dagger(\zeta^*)$. 

The matrix functions ${\Psi_0}_\pm({\zeta})$ (\ref{10}) 
solve the following regular Riemann-Hilbert problem:
\begin{equation}
{\Psi_0}_-^{-1} {\Psi_0}_+ =\Gamma e^{i{\zeta}x\sigma_3}G 
e^{-i{\zeta}x\sigma_3} \Gamma^{-1},
\quad {\rm Im}{\zeta}=0,
\label{13}\end{equation}
where $\Psi_{0\pm}\rightarrow{I}$ as ${\zeta}\rightarrow\infty$.
The solution to the regular Riemann-Hilbert problem is 
unique due to the normalisation condition at infinity.

The spectral data defining a unique solution to the Riemann-Hilbert problem 
consists of two parts: the discrete set of $\zeta_j$ and $|p_j\rangle$ 
($j=1,...,N$) and the continuous data $g(\zeta)=G_{12}(\zeta)$.  
The pure soliton solutions arise from the 
Riemann-Hilbert problem with zeros provided $g({\zeta})=0$, i.e., 
${\Psi_0}_\pm({\zeta})={I}$. 

To derive the coordinate dependence of the discrete data we note that 
$\det\Psi_\pm(\zeta)$ is $t$-independent, hence $(\zeta_j)_t =0$. The 
coordinate dependence of $\mid p_j\rangle$ is obtained 
 by the differentiation of the first relation in (\ref{12}) and using 
equations (\ref{2a}), (\ref{2b}) and (\ref{12}):
\begin{equation}
\mid p_j\rangle_x=i{\zeta}_j\sigma_3\mid p_j\rangle,\quad 
\mid p_j\rangle_t=-2i{\zeta}^2_j\sigma_3\mid p_j\rangle.
\label{14}\end{equation}
Hence
\begin{equation}
\mid p_j\rangle=\exp(f_j\sigma_3)\mid p^{(0)}_j\rangle=
\left(\begin{array}{c}\exp(a_j+i{\theta}_j+f_j)\\[3pt] e^{-f_j}\end{array}
\right),
\label{15}\end{equation}
where $f_j=i{\zeta}_jx-2i{\zeta}^2_jt$ and we have defined 
$\exp(a_j+i{\theta}_j)=(p^{(0)}_j)_1/(p^{(0)}_j)_2$, with $a_j$ 
and  ${\theta}_j$ real constants. 

In what follows we will need the one-soliton dressing matrix. 
For $N=1$ equations (\ref{11}) and 
(\ref{15})  yield
\begin{equation} \Gamma({\zeta})=\frac{{\rm sech} 
z}{2({\zeta}-\zeta^*_1)} 
\left(\begin{array}{cc}({\zeta}-{\zeta}_1)e^z+({\zeta}-\zeta^*_1) 
e^{-z}&-2i\eta e^{i{{\varphi}}}\\[3pt] -2i\eta e^{-i{{\varphi}}}& 
({\zeta}-\zeta^*_1)e^z+({\zeta}-{\zeta}_1)e^{-z} \end{array}\right),
\label{16}\end{equation}
where  we have defined 
\[
z=a+2{\rm Re}f_1\equiv -2\eta( x - x_0),
\]
\begin{equation}
\label{17}\end{equation}
\[
{{\varphi}}={\theta}+2{\rm Im}f_1\equiv -\frac{\xi}{\eta}z + \varphi_0;
\]
decomposed ${\zeta}_1=\xi+i\eta$, and denoted $a_1\equiv a$ and 
${\theta}_1\equiv{\theta}$. In (\ref{17}) we have introduced the  position 
$x_0$ and the core phase   $\varphi_0$ of the soliton, via 
\begin{equation}
x_0\equiv \frac{1}{2\eta}z\biggr|_{x=0}=\frac{1}{2\eta}\left(a+8\eta\xi 
t\right),\quad \varphi_0\equiv
\varphi\biggr|_{x=x_0}={\theta}+2\xi x_0 + 4(\eta^2 - \xi^2) t.
\label{x0phi0}\end{equation}
Finally, the one-soliton solution to the (unperturbed) NLS equation 
parametrised by the Riemann-Hilbert data reads 
\begin{equation}
q_s(x,t)=-2\lim_{\;{\zeta}\rightarrow\infty}{\zeta}\Gamma_{12}({\zeta})=2i\eta 
e^{i{{\varphi}}}
{\rm sech}z.
\label{18}\end{equation}
Here $2\eta$, $4\xi$ and $a$  give, respectively,
the amplitude, velocity and initial position  of the 
soliton. $\theta$ is the initial phase at the point $x=0$:
$\theta=  \varphi \bigr|_{x=t=0}$.


\subsection{Evolution of the spectral data in the perturbed NLS equation}
\label{eq}
When ${\cal R}\ne0$ in equation (\ref{1}), the evolution of the associated 
spectral data becomes nonlinear and complicated. Here our analysis 
follows the lines of Refs.~\cite{7,8,9,11,14,16,18}. To distinguish between 
the integrable and perturbation-induced $t$-dependence, we use the 
``variational derivative" notation. For instance,  the perturbation 
 (\ref{pert}) can be written as 
\[
{\cal R}=he^{2i\Omega_0 t}{q^*}-i\gamma q \equiv i\frac{\delta q}{ 
\delta t}. \]
Introducing an off-diagonal matrix 
\begin{equation} 
R=\left(\begin{array}{cc}0&{\cal R}\\-{\cal R}^*&0\end{array}\right)
\label{19}\end{equation} 
and the linear  functional of the perturbation
\begin{equation}
\Upsilon(\pm\infty,x;{\zeta})=\int\limits_{\pm\infty}^x\,{\rm d}x 
e^{-i{\zeta}x\sigma_3}\Psi_+^{-1}(x,{\zeta})R(x)\Psi_+(x,{\zeta})e^{i{\zeta}x\sigma_3},
\label{upsilon}\end{equation}
we can write the variational derivative of  $\Psi_+$ in the following form
(see \cite{18} for details):
\begin{equation}
\frac{\delta\Psi_+(x,{\zeta})}{\delta 
t}=\Psi_+(x,{\zeta})e^{i{\zeta}x\sigma_3}
\Pi(x,{\zeta})e^{-i{\zeta}x\sigma_3},
\label{21}\end{equation}
where
\[
\Pi(x,{\zeta})=\left(\begin{array}{cc}
-\Upsilon_{11}(x,\infty;{\zeta})&\Upsilon_{12}
(-\infty,x;{\zeta})\\
-\Upsilon_{21}(x,\infty;{\zeta})&\Upsilon_{22}
(-\infty,x;{\zeta})\end{array}\right).
\]
(In (\ref{upsilon}), (\ref{21}) and below we are omitting the explicit 
$t$-dependence for notational convenience.)
The evolution functional $\Pi(x,{\zeta})$ is meromorphic 
in the upper half-plane of  ${\zeta}$ 
and has simple poles at zeros of  $\det\Psi_+({\zeta})$ (which are 
assumed to be simple). The r.h.s. of (\ref{21}) describes the 
perturbation-induced
evolution of  $\Psi_+$ and should be added to the r.h.s. of (\ref{2b}). In 
view of the involution $\Psi^\dagger({\zeta})=\Psi^{-1}(\zeta^*)$, the 
evolution of  $\Psi^{-1}_-$ is given by the
Hermitian conjugate of equation (\ref{21}):
\begin{equation}
\frac{\delta\Psi_-^{-1}(x,{\zeta})}{\delta t}
=e^{i{\zeta}x\sigma_3}\Pi^\dagger(x,\zeta^*)
e^{-i{\zeta}x\sigma_3}\Psi_-^{-1}(x,{\zeta}).
\label{22}\end{equation}   

Relations (\ref{21})-(\ref{22}) yield equations for the 
perturbation-induced evolution of the complete set of the spectral data 
$\{g({\zeta});{\zeta}_j,a_j,{\theta}_j (j=1,\ldots, N)\}$:
\begin{equation}
\frac{{\rm d}{\zeta}_j}{{\rm d}t}=-{\rm res}
\Upsilon_{22}({\zeta}_j),
\label{26a}\end{equation}
\begin{equation}
\frac{{\rm d}}{{\rm d}t}(a_j+i{\theta}_j)
=\Upsilon^{\rm (reg)}_{22}({\zeta}_j)-\exp\left(4i\int\limits^t_0\,{\rm d}t
{\zeta}_j^2-a_j-i{\theta}_j\right)\Upsilon^{\rm (reg)}_{12}({\zeta}_j),
\label{26b}\end{equation}
\begin{equation}
\frac{{\rm d}g({\zeta})}{{\rm d}t}
=-4i{\zeta}^2g({\zeta})  + \Upsilon_{12}({\zeta}) +\Upsilon_{22}({\zeta}) 
g({\zeta}),
\label{26c}\end{equation}
where $\Upsilon({\zeta})=\Upsilon(-\infty,\infty;{\zeta})$ and
\[
\Upsilon^{\rm (reg)}({\zeta_j})=\lim_{\zeta\to\zeta_j}
\left\{\Upsilon({\zeta})-({\zeta}-{\zeta}_j)^{-1}{\rm res} 
\Upsilon({\zeta}_j)\right\} 
\]
is the regular part of  $\Upsilon({\zeta})$ at the pole   
${\zeta}={\zeta}_j$. In the derivation  of (\ref{26a})-(\ref{26c}) we used 
equations (\ref{9}), (\ref{12}), (\ref{15}) and the identities 
\begin{equation}
{\rm res}\,\Pi(x,{\zeta}_j)\mid{\tilde{p}_j}\rangle =
-\frac{{\rm d}{\zeta}_j}{{\rm d}t}\mid{\tilde{p}_j}\rangle ,
\quad \Psi_+(x,{\zeta}_j)e^{i{\zeta}_jx\sigma_3}{\rm 
res}\,\Pi(x,{\zeta}_j)=0,
\label{23}\end{equation}   
where ${\rm res}\,\Pi({\zeta}_j)$ denotes the residue of  $\Pi({\zeta})$ at  
${\zeta}_j$ and $\mid{\tilde{p}_j}\rangle 
=\exp(-i{\zeta}_jx\sigma_3)\mid{p_j}(x)\rangle$.

Equations (\ref{26a})-(\ref{26c}) are highly nonlinear since 
the right-hand sides are dependent on the function $\Psi_+(\zeta)$ which is 
itself to be constructed from the spectral data. However, these equations 
can be simplified by expanding in powers of a suitably chosen small 
parameter. In section 4 we will  make a choice of the small parameter that 
will allow us to develop a rigorous approach to the soliton-radiation 
interaction. For completeness of the presentation, however, we
consider the standard  adiabatic approximation first. 

\section{Adiabatic approximation}
\label{adiabat}
In this section we summarise the main points  of the adiabatic 
approach and establish their connection to some facts about the 
parametrically driven, damped NLS equation, available in literature. 
Consider equation (\ref{1}) with ${\cal R}$ as 
in (\ref{pert}), and assume that $h$ and $\gamma$ are small. In the 
adiabatic approximation the soliton solution is assumed to be given 
simply by the unperturbed NLS soliton (\ref{18})  with the parameters 
$\xi$, $\eta$, $a$, and  $\theta$ being slowly changing functions of time. 
We now derive and discuss the adiabatic equations for the soliton 
parameters.

The matrix $\Upsilon({\zeta})\equiv\Upsilon(-\infty,\infty;{\zeta})$,
the key element of the perturbation theory defined in (\ref{upsilon}),  can 
be easily computed for the NLS soliton with the $t$-dependent parameters. In 
this case $\Psi_{0+}=I$ and $\Psi_+({\zeta})\equiv\Gamma({\zeta})$. Using 
(\ref{16}) we get \begin{equation}
\Upsilon_{22}({\zeta})=\frac{i}{4}\int\limits_{-\infty}^\infty{\rm  d}z\,
{\rm 
sech}^2z\left(\frac{e^z}{{\zeta}-{\zeta}_1}+\frac{e^{-z}}{{\zeta}
-\zeta^*_1}\right)
\Bigl[{\cal R}_0(z)+{\cal R}^*_0(-z)\Bigr]
\label{27}\end{equation}
and
\[
\Upsilon_{12}({\zeta})=\frac{i}{4}\int\limits_{-\infty}^\infty{\rm  d}z
\exp(-2i{\zeta}x+i{{\varphi}}){\rm sech}^2z\Biggl\{\left[
\frac{({\zeta}-\zeta^*_1)e^{2z}}{2i\eta({\zeta}-{\zeta}_1)}
+\frac{({\zeta}-{\zeta}_1)e^{-2z}}{2i\eta({\zeta}-\zeta^*_1)}
+\frac{1}{i\eta}\right]{\cal R}_0(z)
\]
\begin{equation}
+\frac{2i\eta  
{\cal R}^*_0(z)}{({\zeta}-{\zeta}_1)({\zeta}-\zeta^*_1)}\Biggr\}. 
\label{28}\end{equation} 
In equations (\ref{27}) and (\ref{28}) we have omitted, for notational 
convenience, the explicit $t$-dependence and defined 
\begin{equation}
{\cal R}_0=e^{-i{{\varphi}}}{\cal R}. 
\end{equation}
The variables $z$ and  $\varphi$ are defined as 
in (\ref{17}) where this time, we need to  take into account the 
$t$-dependence of the zero $\zeta_1= \xi+ i\eta $.
In this case the integration of (\ref{14}) gives equation (\ref{15}) with
\begin{equation}
f_1=i{\zeta}_1x-2i\int^t_0{\rm  d}t\, {\zeta}^2_1,
\label{f1}\end{equation}
whence the core phase and position of the soliton are found to be 
\begin{equation}
\varphi_0=\theta + 2\xi x_0 + 4\int\limits^t_0{\rm  d}t\left(\eta^2 - 
\xi^2\right),\quad
x_0=\frac{1}{2\eta}\left(a+8\int\limits^t_0{\rm  d}t\xi\eta\right). 
\label{phaspoz}\end{equation} 
Inserting  (\ref{27}) and (\ref{28}) into 
(\ref{26a})  and (\ref{26b}) (with $j=1$), gives
\begin{equation}
\frac{{\rm  d}}{{\rm  d}t}(\xi+i\eta)=-\frac{i}{4}\int\limits_{-\infty}^\infty
{\rm  d}ze^z{\rm sech}^2z\Bigl[{\cal R}_0(z)+{\cal R}^*_0(-z)\Bigr]
\label{30}\end{equation}
and
\begin{equation} 
\frac{{\rm  d}}{{\rm  d}t}(a+i{\theta})=
-\frac{i}{\eta}\Biggl(a+8\int\limits^t_0{\rm  d}t\,\xi\eta\Biggr)
\frac{{\rm  d}{\zeta}_1}{{\rm  d}t}
-\frac{1}{4\eta}\int
\limits_{-\infty}^\infty{\rm  d}z\,{\rm sech}^2z
\Bigl(\cosh z-ze^z\Bigr)
\Bigl[{\cal R}_0(z)-{\cal R}^*_0(-z)\Bigr],
\label{31}\end{equation}
where, as in the previous section, ${\zeta}_1=\xi+i\eta$, and we have used 
 the identity \[
-2i{\zeta}_1x+i{{\varphi}}=i{\theta}+a-z-4i\int^t_0{\rm  d}t\,{\zeta}_1^2.
\]
(This follows from (\ref{17}) and (\ref{f1}).)

Let ${\cal R}_{0s}$ stand for the value of the quantity ${\cal R}_0$ 
calculated on the soliton, i.e., with the phase and position parameters 
given by (\ref{phaspoz}). Equations  (\ref{30}) and  (\ref{31}) with
${\cal R}_0={\cal R}_{0s}$ constitute the set of the adiabatic equations for 
the soliton parameters.  These equations are equivalent to those derived by 
Karpman and Maslov \cite{8}.

For the parametrically driven, damped NLS equation (\ref{1})-(\ref{pert}) 
the quantity ${\cal R}_{0 s}$ equals
\begin{equation}
{{\cal R}_0}_s(z)=e^{-i{{\varphi}}}\Bigl(he^{2i\Omega_0 t}q^*_s-i\gamma 
q_s\Bigr) =2\eta\left\{\gamma-ih\exp\Bigl[2i(\Omega_0 
t-{{\varphi}})\Bigr]\right\}{\rm sech}z.
\label{32}\end{equation}
Substituting this into (\ref{30}) and  (\ref{31}) we obtain, after some 
algebra: \begin{equation}
\frac{{\rm  d}\eta}{{\rm  d}t}=-\frac{2\pi{h}\xi}
{\sinh(\pi\xi/\eta)}\sin(2\Omega_0 t-2\varphi_0) - 2\gamma\eta,
\label{33a}\end{equation}
\begin{equation}
\frac{{\rm  d}\xi}{{\rm  d}t}=\frac{2\pi{h}\xi^2/\eta}
{\sinh({\pi\xi}/{\eta})}\sin(2\Omega_0 t-2\varphi_0),
\label{33b}\end{equation}
\[
\frac{{\rm  d}}{{\rm  d}t}(a+i{\theta})=
\frac{\pi{i}h\xi/\eta}{\sinh({\pi\xi}/{\eta})}
\Biggl\{\cos(2\Omega_0 t-2\varphi_0)
\left[\frac{i\eta}{\xi}-\pi\coth\biggl(\frac{\pi\xi}{\eta}\biggr)
\biggl(i-\frac{\xi}{\eta}\biggr)\right]\Biggr\}
\]
\begin{equation}
+2x_0\frac{{\rm  d}}{{\rm  d}t}(\eta - i \xi).
\label{34}\end{equation}
Equations  (\ref{33a})  and (\ref{33b}) 
imply ${\rm  d}(\eta\xi)/{\rm  d}t=-2\gamma \eta\xi$, i.e the 
quantity $\eta\xi$, proportional to the momentum of the soliton \cite{6},
has to decay to zero as $t\to\infty$.   
Taking this into account we will restrict ourselves to the nonpropagating 
soliton: $\xi=0$. We can also choose $a=0$ so that the soliton is placed at 
the origin: $x_0=0$, see equation (\ref{phaspoz}). Thus, sending in 
(\ref{33a}) and (\ref{34})  $\xi\to0$ and making use of the identity \[ 
\lim_{\xi\rightarrow0}\frac{\pi\xi/\eta}{\sinh\Bigl({\pi\xi}/{\eta}\Bigr)} 
\left[\frac{i\eta}{\xi}-\pi\coth\biggl(\frac{\pi\xi}{\eta}\biggr) 
\biggl(i-\frac{\xi}{\eta}\biggr)\right]=1, \] we arrive at a closed system 
of equations for the soliton's amplitude and phase. This can be presented  
in the following convenient form: \begin{eqnarray} {\dot \eta}= -2 \eta 
(\gamma + h \sin \Phi), \label{a1}\\ {\dot \Phi}= 2 \Omega_0- 8 \eta^2- 2 h 
\cos \Phi, \label{a2}
\end{eqnarray}
where
\begin{equation}
\Phi/2 = \Omega_0 t -\varphi_0
\label{Z}\end{equation} 
is the difference between the phase of the driver and the core phase of the 
soliton.

These equations were first derived  in  \cite{BKM}  and \cite{Fauve}
(within a different approach, though).
The two first-order equations (\ref{a1})-(\ref{a2}) can be rewritten as a 
single second-order equation for $\Phi$:
\begin{equation}
{\ddot \Phi}+2\Bigl(2\gamma+h\sin{\Phi}\Bigr)
{\dot \Phi}- 8 \Bigl(\Omega_0 -h\cos{\Phi}\Bigr)
\Bigl(\gamma+h\sin{\Phi}\Bigr)=0,
\label{Z1}
\end{equation}
which in some situations is more amenable for  analysis.
Equations (\ref{a1})-(\ref{a2}) have two fixed points $(\eta_{\pm}$, 
$\Phi_{\pm})$  which correspond to stationary soliton solutions 
(\ref{18}) with $\eta=\eta_{\pm}$ and $\Phi=\Phi_{\pm}$.   Here 
\begin{equation} 2\eta_{\pm}= \sqrt{\Omega_0 \pm H},
\quad
{\Phi_+}=  -\pi+\arcsin\left(\frac{\gamma}{h}\right)
, \quad {\Phi_-}=
-\arcsin\left(\frac{\gamma}{h}\right)
\label{param}\end{equation}
and
\[
H=\sqrt{h^2-\gamma^2}.
\]
Although obtained in the adiabatic approximation, these two solitons turn 
out to be exact solutions of the parametrically driven damped NLS 
\cite{F1,M5}. 

Linearising   equation (\ref{Z1}) in the small 
perturbation $\phi= \Phi-\Phi_{\pm}$, and making use of relations
\begin{equation}
h\sin(\Phi_\pm+\phi)=-\gamma\mp H{\phi}+{\cal O}(\phi^2),
\quad
h\cos(\Phi_\pm+\phi)=\mp H+\gamma{\phi}+{\cal O}(\phi^2),
\label{approxi}
\end{equation}
yields the equation of  damped linear oscillator:
\[
{\ddot \phi} +2\gamma
{\dot \phi} \pm 8 H\Bigl( \Omega_0 +  H \Bigr)\phi=0.
\]
From here one readily concludes \cite{BKM,Fauve} that the $q_+$-soliton 
(i.e. the soliton (\ref{18}) with $\eta=\eta_+$ and $\Phi=\Phi_+$) is 
adiabatically stable and,
when excited, exhibits decaying oscillations at the bare frequency
$\Omega_s=\sqrt{8H (\Omega_0+H)}$.
The $q_-$-soliton (for which $\eta=\eta_-$ and $\Phi=\Phi_-$) is 
adiabatically unstable, and this of course
implies that it is unstable within the full partial differential equation.
(This is corroborated by the stability analysis of the full
equation, see \cite{M5}.)
For this reason  we disregard the $q_-$ soliton in what follows
 and focus entirely on the $q_+$; from now on the ``parametrically driven damped soliton"
will always mean the soliton $q_+$.

It is interesting to note that the adiabatic equations
(\ref{a1})-(\ref{a2}) have another solution that admits a simple  
interpretation. It is given by $\eta=0$ and ${\dot \Phi}=2 (\Omega_0-h \cos 
\Phi)$ and corresponds to the flat solution $q=0$ of the full 
damped-driven NLS. It will reappear in our analysis of the soliton-radiation 
interaction below.

The main shortcoming of the adiabatic approximation is in that it ignores 
the soliton-radiation interaction. As a result, it 
is unable to reproduce many features of soliton's dynamics even for fairly 
small perturbations. In particular, the adiabatic approach does not capture 
the oscillatory instability of the $q_+$ soliton which sets in as the 
driving strength exceeds a certain --- rather low --- threshold \cite{M5}. 
(For example, for $\gamma=0$ this threshold is at $h=h_c(0)=0.064$.) Neither 
is it capable of reproducing 
secondary bifurcations and chaotic dynamics of the 
soliton. In what follows we go beyond the adiabatic approximation and take 
the soliton-radiation coupling into account.

\section{Evolution of the spectral data}
\label{second}
\subsection{The essence of our approach}
\label{essence}
As we mentioned in section 2, the soliton of the unperturbed, ``pure", NLS
corresponds to a single zero of the Riemann-Hilbert problem
associated with this equation. Our approach to the parametricaly driven, 
damped NLS (\ref{1})-(\ref{pert}) is based on the fact that it can be  cast 
in the form 
\begin{equation} i u _t+ u _{xx}+2|{ u }|^2{ u }=he^{2i\Omega t} 
u^* -( H +i\gamma){ u }, 
\label{equ1}\end{equation}
where
\begin{equation}
u(x,t)=e^{iH t} q(x,t),
\label{Ht}\end{equation}
and $\Omega= \Omega_0 + H $. The key property of the new formulation is that 
both the left and the right-hand side of equation (\ref{equ1}) are equal to 
zero for $u$ equal to \begin{equation}
 u_+(x,t)=2i\eta_+ \frac{\exp \left\{  4i\eta_+^2t - (i/2) \Phi_+\right\}}
{\cosh\left( 2\eta_+ x \right)},
\label{sol}
\end{equation}
where $\Phi_+$ and $\eta_+$ are as in
(\ref{param}). (In particular, $2\eta_+= \sqrt{\Omega}$.)
Hence $u_+$  coincides --- {\it exactly} ---
 with the soliton of the 
unperturbed NLS equation with a particular amplitude and phase  (selected 
by the parameters of the perturbation). Therefore, this solution of the 
perturbed equation is also associated with a single zero of the 
Riemann-Hilbert problem underlying the unperturbed, integrable, NLS. It is 
important to emphasise that this correspondence is valid for {\it 
arbitrarily large\/} values of $h$ and $\gamma$ (where $h$ only has to be  
greater than $\gamma$ so that the soliton (\ref{sol}), (\ref{param}) 
exists).

Rescaling $t,x, q,h$ and $\gamma$ we can always arrange that $\Omega_0=1$  
in equation (\ref{pert}). Next, we have already seen that the motionless 
soliton is a solution of the adiabatic equations. It is not difficult to 
realise that, in our case, even when the radiations are taken into account, 
an initially quiescent soliton will remain nonpropagating at all times. 
This follows from the evolution equations for the spectral data  
$\{\eta, \xi,x_0,\varphi_0; g(\zeta)\}$ for the damped-driven NLS 
(\ref{1})-(\ref{pert}). These evolution equations, linearised in $\varphi_0$ 
 and $g(\zeta)$, can be easily shown to be compatible with the constraint 
\begin{equation} x_0=0, \quad \xi=0,\quad g(-\zeta) = g(\zeta).
\label{nonpr}\end{equation}
Consequently, in this paper we confine ourselves to the internal dynamics of 
the {\it nonpropagating} soliton (placed at the origin for convenience) and 
its radiation. The corresponding solution of equation (\ref{equ1}) is given 
by an even function of $x$:
\begin{equation}
 u (x,t)=u_s(x,t) + u _{r}(x,t),
\label{solution}
\end{equation}
where $u_s(x,t)$ has the form of the
motionless  soliton of the ``pure" NLS, located at the origin,
with the time-dependent amplitude and phase:
\begin{equation}
u_s(x,t)=2i \eta e^{i\varphi_0}{\rm sech}z. 
\label{vari_sol} \end{equation}
Here $\eta=\eta(t)$,
$z=-2\eta(t) x$, and 
$\varphi_0$ is given by equation (\ref{phaspoz}) where we 
only need to set $\xi=0$: 
\[
\varphi_0(t)= \theta(t) + 4\int\limits_0^t\eta^2{\rm d}t.
\] 
Equivalently, the soliton can be written in the form
\[
u_s(x,t)=2i \eta \frac{\exp \left\{ i \Omega t - i \Phi/2
\right\}}{\cosh
z},
\]
where the variable $\Phi$ is defined by 
\begin{equation} \Phi/2 = 
\Omega t - \varphi_0. \label{def_Phi}
\end{equation}
The definition  (\ref{def_Phi}) is equivalent to (\ref{Z}); in both  cases 
$\Phi/2$ is the difference between the phase of the driver and the 
core phase of 
the soliton. (Here we should alert the reader
to the fact  that, since $q$ and $u$ are different by the factor
  (\ref{Ht}), 
the core phases  $\varphi_0(t)$ in (\ref{def_Phi}) and 
 in (\ref{Z}) are different by $Ht$.)

The  second term in (\ref{solution}) accounts for the radiation waves. As in 
Ref.~\cite{add1}, in our derivation of the evolution equations for the 
spectral data we will retain only terms linear in radiation. 
Hence it is sufficient to solve the {\it linearised} version of the regular 
Riemann-Hilbert problem~(\ref{13}) to obtain the radiation part of the 
solution~(\ref{solution}). The linearisation of the Riemann-Hilbert problem 
produces  the Plemelj jump problem:  
\begin{equation}
{\Psi_0}_+ - {\Psi_0}_- = \Gamma e^{i{\zeta}x\sigma_3}G 
e^{-i{\zeta}x\sigma_3} \Gamma^{-1}, \quad {\rm Im\/} \zeta=0.
\label{linRH}\end{equation}
Taking into account the normalisation condition $\Psi_0\to I$ as $\zeta\to 
\infty$,  we obtain from (\ref{linRH}): 	
\begin{equation} 
{\Psi_0}_\pm({\zeta})={I}+\frac{1}{2\pi i}\int\limits^{\infty}_{-\infty} 
\frac{{\rm  d}l}{l-{\zeta}}\Gamma(l)\left(\begin{array}{cc} 0&e^{2ilx}g(l)\\ 
e^{-2ilx}g^*(l)&0\end{array}\right)	 \Gamma^{-1}(l). 
\label{40}\end{equation} 
(Here the sign $+$ respectively $-$ indicates
that  $\zeta$ lies in the upper  respectively lower  half-plane.) 
The radiation part of the solution (\ref{solution}) is now given 
 by equation 
(\ref{qlim}):
 \begin{equation} 
{u}_{r}=-2\lim_{{\zeta}\rightarrow\infty}{\zeta}({\Psi_0}_+)_{12} 
=\frac{1}{\pi i}\int\limits_{-\infty}^\infty{\rm  d}{\zeta}\left[ 
\Gamma_{11}(\Gamma^{-1}){}_{22} \, 
e^{2i{\zeta}x}g+\Gamma_{12}(\Gamma^{-1}){}_{12} \, 
e^{-2i{\zeta}x}{g^*}\right]. 
\label{ur1}\end{equation}
Finally, substituting the one-soliton dressing matrix (\ref{16})
into equation (\ref{ur1}), introducing the notation
\begin{equation}
k=\frac{{\zeta}}{\eta} 
\label{spectral}
\end{equation} 
and defining the radiation amplitude $b(k)$ by 
\begin{equation}
b(k)=\exp(-i\varphi_0)g(\eta k),
\label{gandb}\end{equation}
we arrive at the formula for the linear radiation:
\begin{equation}
{u}_{r}(x,t)=\frac{\eta e^{i{{\varphi_0}}}}{\pi 
i}\int\limits^{\infty}_{-\infty} \frac{{\rm  d}k}{k^2+1}\left[
\left(k- i\tanh\!z\right)^2e^{-ik z}b(k)  + {\rm  sech}^2z 
e^{ik z}b^*(k)\right]
\label{qr}\end{equation}
--- in exact agreement with the corresponding result in Ref.~\cite{8}.

\subsection{A closed system for the evolution of the spectral data}

Since the r.h.s. of the perturbed  NLS (\ref{equ1}) 
is linear in $u$, the decomposition of the solution 
into  the  soliton and radiation parts, equation (\ref{solution}),   
induces the corresponding decomposition of the perturbation:
\[
{\cal R}={\cal R}(u_s)+{\cal R}(u_r)\equiv{\cal R}_s+{\cal R}_r.
\]
The perturbation matrix (\ref{19}) splits accordingly: $R=R_s +R_r$. 
Substituting for $u_s$ and $u_r$ from (\ref{vari_sol}) and (\ref{qr}), we 
get  
\[ {\cal R}_{s}=(R_s)_{12}=he^{2i{\Omega t}}u^*_{s}-({H}+i\gamma)u_{s} 
=\left(-i{H}+\gamma-ihe^{i{\Phi}}\right)2\eta e^{i{{\varphi_0}}}{{\rm  
sech}} z 
\] 
and
\[
{\cal R}_{r}=(R_r)_{12}=he^{2i{\Omega t}}u^*_{r}-({H}+i\gamma)u_{r},
\]
respectively.
Discarding terms higher than linear in $b$ in the expansion  of the 
functional  (\ref{upsilon}) yields 
\begin{equation} 
\Upsilon(k)=\int\limits_{-\infty}^\infty\,{\rm d}x \exp(-ik\eta x 
\sigma_3) \Gamma^{-1}(k\eta) \Bigl\{R_s + R_r + 
[R_s,{\Psi}_{0+}(k\eta)]\Bigr\}\Gamma(k\eta) \exp(ik\eta x \sigma_3), 
\label{ups1}\end{equation} where $\Gamma(\zeta)$ is the one-soliton dressing 
matrix (\ref{16}) and ${\Psi}_{0+}(\zeta)$ is given by equation  (\ref{40}). 

In order to obtain the  evolution equations for the 
spectral data one has to evaluate the integrals in the expression 
(\ref{ups1}); substitute the resulting matrix $\Upsilon$ in equations 
(\ref{26a})-(\ref{26c}) and use the simplifying conditions (\ref{nonpr}) 
for the nonpropagating soliton. (As we already mentioned, these conditions 
are compatible with the evolution equations for the spectral data.) It is 
convenient to present the final result in terms of $\Phi$;
 $w=8\eta^2$; and 
the real and imaginary parts of the radiation amplitude $b(k)=b_R+ib_I$.  
After tedious but otherwise straightforward  calculations one arrives 
at the following system: 
\begin{equation} 
{\dot \Phi} = 2(1 - h\cos{\Phi}) - w, 
\label{sys1}
\end{equation} 
\begin{equation} 
{\dot w} = -\Biggl\{ 
(\gamma+h\sin{\Phi})\left[4-\int\limits^{\infty}_{-\infty} {\rm d}k {\rm 
sech}\left(\frac{\pi k}{2}\right)b_R(k)\right] 
-2h\cos{\Phi}\int\limits^{\infty}_{-\infty}{\rm d}k {\rm 
sech}\left(\frac{\pi k}{2}\right)b_I(k) \Biggr\}w, 
\label{sys2}\end{equation} 
\[
\frac{\partial b_R}{\partial t} = \left[\frac{w}{2}(1+k^2)+2h\cos{\Phi}
(I+\hat{\cal K})\right]b_I - 
(\gamma+h\sin{\Phi})\left[b_R+2k\frac{\partial b_R}{\partial 
k}+\left(\hat{\cal M}_+ -\hat{\cal M}_-\right)b_R\right]
\]
\begin{equation}
-(\gamma+h\sin{\Phi})\pi{\rm sech}\left(\frac{\pi k}{2}\right),
\label{sys3}\end{equation}
\begin{equation}
\frac{\partial b_I}{\partial t} = -\frac{w}{2}(1+k^2)b_R 
-2\gamma b_I + (\gamma+h\sin{\Phi})\left[b_I - 2k\frac{\partial 
b_I}{\partial k}-\left(\hat{\cal M}_+ +\hat{\cal M}_-\right)b_I\right].
\label{sys4}\end{equation}
Here the operators $\hat{\cal K}$ and $\hat{\cal M}_\pm$ are defined on
even functions: 
\[ \hat{\cal K} f(k) = 
\int\limits^{\infty}_{-\infty}{\rm 
d}l\frac{(l-k)}{\sinh[\pi(l-k)/2]}\frac{(1-lk)}{(1+k^2)(1+l^2)}f(l), 
\] \[
\hat{\cal M}_\pm f(k) = \frac{1}{2}\int\limits^{\infty}_{-\infty}{\rm 
d}l\frac{(l\pm k)}{\sinh[\pi(l-k)/2]}\frac{(l^2+k^2+2)}{(1+k^2)(1+l^2)}f(l), 
\] where the singular integral in the expression for $\hat{\cal M}_+$ should 
be understood in the  sense of the Cauchy principal value. 

In equations 
 (\ref{sys3}) and  (\ref{sys4}), the notation 
 $\partial b/\partial t$ 
 is meant to indicate that the derivatives are taken for fixed $k$.
  (On the contrary, writing ${\dot b}$ would mean $\partial b/\partial t
  + {\dot k} \partial b/ \partial k$ with $k(t)$ as in (\ref{spectral}),
  $k=\zeta/\eta(t)$.) We are using partials here for  later computational
  convenience. 

It is worth re-emphasising that equations (\ref{sys1})-(\ref{sys4})
are valid for arbitrarily large $h$ and $\gamma$. The only approximation
we made in their derivation, was to drop terms of order higher than
linear in $b$.

Letting $b_R=b_I=0$  reduces equations (\ref{sys1})-(\ref{sys2}) to the 
adiabatic equations (\ref{a1})-(\ref{a2}).  Like the adiabatic equations, 
the system (\ref{sys1})-(\ref{sys4}) has a fixed point $w=2\Omega=2(1+H)$, 
$\Phi =\Phi_+\equiv -\pi +\arcsin(\gamma/h)$,  $b_R=b_I=0$, which 
corresponds  to the soliton  (\ref{sol})  of equation  
(\ref{equ1}). Another meaningful solution arises by setting
$w=0$ and solving (\ref{sys1}) for $\Phi(t)$;  the 
$b_R$ and $b_I$ are then recovered from the nonhomogeneous linear system
 (\ref{sys3})-(\ref{sys4}). We will return to (a descendant of)
 this solution in
 section 5.  

\subsection{Linearised evolution of the spectral data}
\label{linearised}

Linearising the system (\ref{sys1})-(\ref{sys4})  about the above fixed
point,
one obtains  four first-order linear equations which are equivalent 
to a pair of  second-order equations for  
$\phi\equiv\Phi-\Phi_+$ and $\beta(k)$, where  
\begin{equation} \beta(k) \equiv \frac{2}{\sqrt{\pi}}\frac{b_I(k)}{\sqrt{1+k^2}}. 
\label{beta}\end{equation} 
The second-order system has the form: 
\[
{\ddot \phi}
+2\gamma {\dot \phi}+8 H \Omega{\phi}=2\sqrt{\pi} H \Omega
\int\limits_{-\infty}^\infty {\rm d}{k}\,
\sqrt{1+{k}^2} \,{\rm sech\/} \left(\frac{\pi{k}}{2}\right)\beta({k}),
\]
\begin{equation}
\label{system}\end{equation}
\[
{\ddot \beta}+2\gamma {\dot \beta}+ {\hat{\cal G}}\beta=-2\sqrt{\pi}
H\Omega \sqrt{1+{k}^2} \,{\rm sech\/}\left(\frac{\pi{k}}{2}\right)\phi.
\]
Here  we have  made use of 
relations (\ref{approxi})
and introduced a real symmetric operator ${{\hat {\cal G}}}$:
\begin{equation}
{\hat{\cal G}}= ( 1+\Omega {k}^2)^2 -{H}^2 +2H\Omega{\hat {\cal B}},
\label{opG}
\end{equation}
where
\begin{equation}
{{\hat {\cal B}}}f({{k}}) \equiv \int\limits_{-\infty}^{\infty}{{\rm d}}{l}
\frac{({{k}}-{l})({l}{{k}}-1)} {\sinh[\pi(k-l)/2]}
\frac{f({l})}{\sqrt{ (1+{{k}}^2)(1+{l}^2)}}.
\label{opR}
\end{equation}
It is worth noting here that since the equation for $\phi$ is not uncoupled 
from the equation for $\beta$, the $\phi$ and $\beta({k})$  do not
represent the normal modes of the soliton-radiation system.  This is of 
course a consequence of the nonintegrability of the perturbed nonlinear
Schr\"odinger equation.

The system (\ref{system}) is  exactly equivalent to the
NLS  (\ref{equ1}) linearised about the soliton (\ref{sol}).
In particular, one can readily check that for $H \to 0$,  the oscillation
frequencies of (\ref{system}) reproduce the asymptotic expansions obtained 
in \cite{M5}. Indeed, letting $\phi(t)=e^{(i\omega - \gamma)t}\chi$  and 
$\beta(k;t)= e^{(i\omega - \gamma)t}{\alpha}(k)$ transforms 
(\ref{system}) to an eigenvalue problem \[
8 H \Omega{\chi}-
2\sqrt{\pi} H \Omega
\int\limits_{-\infty}^\infty {\rm d}{k}\,
\sqrt{1+{k}^2} \,
{\rm sech\/} \left(\frac{\pi{k}}{2}\right)
{\alpha}({k})=\mu\chi ,
\]
\begin{equation}
\label{EV_system}
\end{equation}
\[
 2\sqrt{\pi}
H\Omega \sqrt{1+{k}^2} \,{\rm sech\/}
\left(\frac{\pi{k}}{2}\right)\chi 
+{\hat{\cal G}}
{\alpha}(k)= \mu {\alpha}(k),
\]
where $\mu= \omega^2 + \gamma^2$. (Below, we normalise the eigenvector
$\{ \chi, {\alpha}(k) \}$ by
 setting $\chi=1$.) 

A simple perturbation analysis shows that for small $H$ there is an 
eigenvalue $\mu= 8H[1+(\pi^2/6-1)H]+O(H^3)$.
Another discrete eigenvalue detaches from 
the boundary of the continuous spectrum (which extends from 
$\mu=1-H^2$ to infinity): $\mu = 1- 9H^2 +O(H^3)$.
 Both eigenvalues
coincide,
to within terms ${\cal O}(H^3)$, with the corresponding 
 expressions for the eigenfrequencies in the space of 
fields $u(x,t)$ \cite{M5}.

In fact the eigenvalue problem (\ref{EV_system}) is amenable to 
analytical study not only in the $H \to 0$ limit. (This is
the principal  advantage of the analysis in the space of
scattering data over the linearisation in the $u(x,t)$-space
where one has to resort to the help of computer.)
The  radiation can be diagonalised in the basis of eigenfunctions
of the operator ${\hat {\cal G}}$:
\begin{equation}
{\hat {\cal G}} \psi({k})= E \psi({k}).
\label{EV}
\end{equation}
This is equivalent to the following differential  eigenvalue  problem:
\begin{equation}
\left[  \left( 1- \Omega \frac{{\rm d}^2 }{{\rm d} \xi^2} \right)^2 -H^2 -E 
\right] \left( 1 -\frac{{\rm d}^2}{{\rm d} \xi^2} \right)y=
4H \Omega \left[ \frac{{\rm d}}{{\rm d} \xi} \left( {\rm sech}^2 \xi \, 
\frac{{\rm d}y}{{\rm d} \xi} \right) + {\rm sech}^2 \xi \, y \right].
\label{SL}
\end{equation}
Here $y=y(\xi)$ is the Fourier cosine transform of the function
$(1+{k}^2)^{-1/2}\psi({k})$:
\begin{equation}
y(\xi)=\int\limits_{-\infty}^{\infty}
\frac{{{\rm d}}{{k}}}{\sqrt{2\pi}} \cos({{k}}\xi)
\frac{\psi({k})}{\sqrt{1+{k}^2}}.
\label{pfipsi}
\end{equation}
The  continuous spectrum of the operator ${\hat {\cal G}}$
occupies the  semiaxis  $E \ge {1-H^2}$ while discrete eigenvalues satisfy 
$E_n < 1-H^2$. A more accurate bound on discrete eigenvalues is established 
in the Appendix:
\begin{equation}
(1-H^2) \left(1-\sqrt{\frac{5}{2}} \frac{H \Omega}{1-H^2}
\right) < E_n < 1-H^2.
\label{window}
\end{equation}
Equation (\ref{window}) implies that the operator ${\hat {\cal G}}$
cannot have any discrete eigenvalues for $H=0$. However, there is at least 
one discrete eigenvalue for any positive $H$ (see the Appendix.)

Expanding the radiation amplitudes ${\alpha}({{k}})$ over the
orthonormalised basis of
(even) eigenfunctions
of ${\hat {\cal G}}$:
\[
{\alpha}({{k}})=\sum_n{c}_n\psi_{E_n}({{k}})
+\int\limits_{1-H^2}^\infty
{\rm d}E\,{c}(E)\psi_E({{k}}),
\]
the eigenvalue problem (\ref{EV_system})  is cast in the form
\begin{equation}
(8H\Omega - \mu)=\sum_{n} {\kappa}(E_n)c_n
+\int\limits_{1-H^2}^\infty{\rm d}E \, {\kappa}(E)c(E),
\label{ch1}
\end{equation}
\begin{equation}
( \mu-E_n)c_n=  {\kappa}(E_n),
\label{ch2}
\end{equation}
\begin{equation}
(\mu -E)c(E) = {\kappa}(E),
\label{ch3}
\end{equation}
where $\kappa$ measures
the coupling  of the soliton and radiation modes:
\begin{eqnarray}
\kappa(E)=2 \sqrt{\pi}
H \Omega\int\limits_{-\infty}^\infty{\rm d} {{k}}
\,\sqrt{1+{{k}}^2}\,
{\rm sech}\left(\frac{\pi {{k}}}{2}\right)\,\psi_E({{k}})
\nonumber \\
=  2\sqrt{2}H \Omega \int\limits_{-\infty}^\infty{\rm d} \xi\,
{{\rm sech}}\,{\xi}\left(1-\frac{{\rm d}^2}{{\rm d}\xi^2}\right)y_E
(\xi).
\label{kappa}
\end{eqnarray}
Here ${\hat {\cal G}}\psi_E({{k}})=E \psi_E({{k}})$;
$\psi({k})$ and $y(\xi)$ are related by the Fourier transform (\ref{pfipsi}).

\begin{center}
\begin{figure}
\psfig{file= 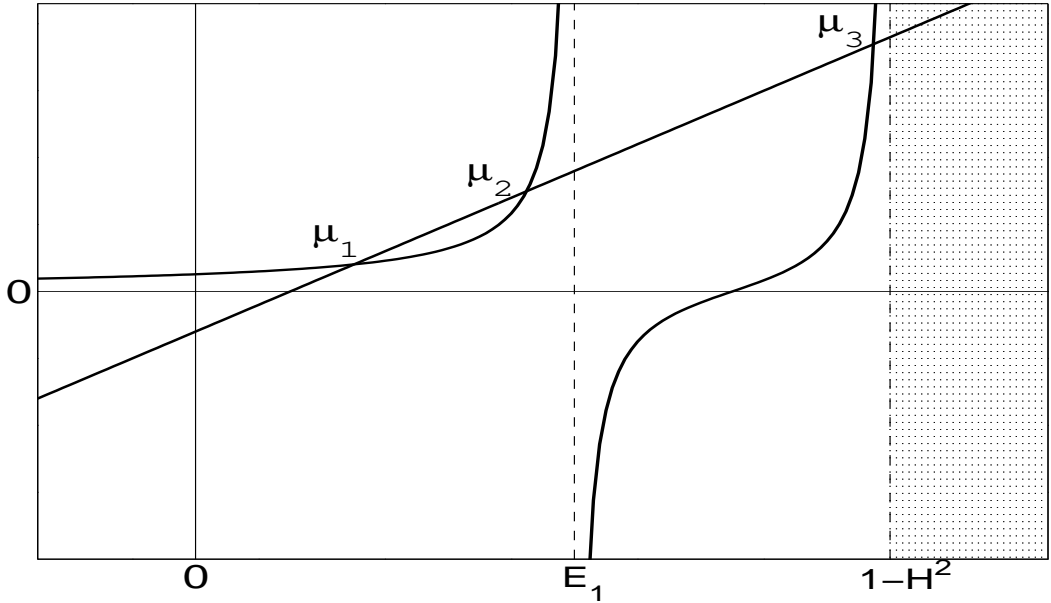,height=0.5\textwidth,width=0.85\textwidth}
\caption{\sf The graphical solution of the characteristic  equation
(\ref{graph}). The solid straight line depicts the function
$y(\mu)=\mu - 8 H \Omega$ and the two solid curves intersecting it
are branches of $y={\frak g}(\mu)$.}
\label{g_omega}
\end{figure}
\end{center}

Solving (\ref{ch2}) and (\ref{ch3}) for $c_n$ and $c(E)$, respectively,
and substituting these into (\ref{ch1}) we arrive at the characteristic
equation of the form
\begin{equation}
\mu -8H\Omega = {\frak g}(\mu),
\label{graph}
\end{equation}
where the function
\begin{equation}
{\frak g}(\mu)= {\frak g}(\mu;H) =
 \sum_n \frac{\kappa^2(E_n)}{E_n-\mu}
+ \int\limits_{1-H^2}^{\infty} {\rm d} E \frac{\kappa^2(E)}
{E- \mu}
\label{g}
\end{equation}
  is defined for
$\mu<1-H^2$.
This is a monotonically growing function for all $\mu$
except points  $\mu=E_n$ where it drops from plus
to minus infinity.  For $\mu < E_1$ the function
${\frak g}(\mu)$ is strictly positive and decays to zero as $\mu \to
-\infty$. (See Fig.\ref{g_omega}).
Roots of equation (\ref{graph}) give discrete frequencies of
the system (\ref{system}), while the 
spectrum of continuous frequencies extends from
$\mu= 1-H^2=1+ \gamma^2-h^2$, to infinity.

Before discussing the roots, we would like  to note that the characteristic 
equation (\ref{graph}) does not contain $\gamma$ explicitly. Having solved 
it for $\mu$ we can recover $\omega$ for {\it any\/} $\gamma$:
$\omega=\sqrt{\mu-\gamma^2}$. This invariance of the eigenvalue problem in 
the space of scattering data is an exact equivalent of the invariance of the 
linearised eigenvalue problem in the space of solutions to the NLS equation 
\cite{M5}.

The graph Fig.\ref{g_omega} shows that the function $y={\frak g}(\mu)$ and the 
straight line $y=\mu - 8H\Omega$ have at least three intersections, $\mu_1$, 
$\mu_2$ and $\mu_3$ such that $\mu_1 \le \mu_2 < E_1 < \mu_3$. As $H$ grows, 
the straight line is shifted down and the function 
${\frak g}(\mu,H)$ also changes. 
In  particular, for small $H$ the coefficients $\kappa^2$ increase in 
proportion to $H^2$ while  the eigenvalue $E_1$ decreases and the edge of 
the continuous spectrum, $1-H^2$, also moves to the left.  Hence for any 
fixed $\mu <E_1$ the corresponding ${\frak g}(\mu)$ grows. As a result, the 
intersection points $\mu_1$ and $\mu_2$ approach each other, then merge for 
some critical value $H=H_{\rm cr}$ and emerge into the complex plane. As $H$ is 
increased past $H_{\rm cr}$, the imaginary parts of $\mu_2=\mu_1^*$ grow and 
eventually the positive imaginary part becomes equal to $\gamma$. This is 
the point of the Hopf bifurcation; for $H$ above this point the soliton 
(\ref{sol}) is unstable.

Conclusions of the above graphical analysis are in exact agreement with the 
behaviour of eigenvalues of the linearisation of  equation (\ref{equ1}) 
observed numerically \cite{M5}. A new feature is the appearance of another
discrete eigenvalue, $\mu_3$, which was not detected in the numerical 
computations of \cite{M5,add2}. One reason for this 
omission could be that the separation  of $\mu_3$ from the continuum  
remains exponentially small for small $H$. Indeed, for $H 
\to 0$ equation (\ref{kappa}) infers that $\kappa(E_1) =4 \pi H^{3/2} 
+O(H^{5/2})$ and $\kappa(1-H^2) = \sqrt{8 \pi} H +O(H^2)$. Assuming 
$\mu_3=1-H^2 -\delta \mu$, where $\delta \mu / H^2 \to 0$ as $H \to 0$, 
equation (\ref{g}) gives ${\frak g}(\mu,H)=-8 \pi H^2 \ln \delta \mu - 2 \pi^2 H +O( 
H^2)$. Substituting into (\ref{graph}), we obtain for the separation $\delta 
\mu$: \begin{equation} \delta \mu  = \exp \left\{ -\frac{1}{8 \pi} 
\frac{1}{H^2} - \frac{2 \pi^2-8}{8 \pi } \frac{1}{H} + C
\right\} \quad \mbox{as} \ H \to 0,
\end{equation}
where $C=C(H)$ remains bounded as $H \to 0$. As $H$ grows, the frequency 
$\mu_3$ remains real and therefore does not give rise to any instabilities 
of the stationary solitons. However it may play a role in the resonance 
phenomena involving oscillating solitons.

We conclude this section with two remarks. First, our linearisation
in the space of scattering data allows to explain the origin of the
oscillatory instability of the soliton. When $H$ is very small,
 the soliton oscillations are virtually uncoupled from the radiation waves.
The former have their frequency close to $\mu =0$ while the
radiations have continuum of frequencies occupying the semiaxis
$\mu \ge 1$. As $H$ is increased, the coupling grows, 
and as a result of that the soliton's
frequency is dragged closer to the continuum while another
local mode is pulled out of the radiation spectrum. Finally, the
two modes merge and the resonance occurs.

Second, we observe that as $k$ grows,
the right-hand side of the bottom
 equation in (\ref{system})  decreases rapidly. We also notice that
  only $\beta(k)$
with small $k$ contribute
significantly to the right-hand side of the top equation.
Consequently, only long radiation waves couple to the soliton. In the
next section we study the effect of the $k=0$ radiation on the soliton's
{\it nonlinear\/} dynamics.

\section{The long-wavelength limit}
\label{third}

In order to study the soliton-radiation interaction at the long-wavelength 
limit, we write the spectral density $b(k)$ in the form
$b({k})=b_{\epsilon}(k)= \epsilon^{-1}\tilde{b}({k}/\epsilon)$,
where $\tilde{b}$ is an even  function of its argument;
integrate equations (\ref{sys3}) and (\ref{sys4}) over 
a small range of $k \in [-\delta k,\delta k]$
and then send $\epsilon \to 0$ in 
(\ref{sys1})-(\ref{sys4}). This allows to derive a  
finite-dimensional system without any {\it ad-hoc \/} cut-off parameters 
(cf. \cite{add1,SPLE,TSFGR}). The resulting four-dimensional system 
comprises equations for the amplitude and phase of  the soliton, and the 
complex amplitude of the ${k}=0$ radiation: \[ {\rho} +i{{\sigma}} = 
\lim_{\epsilon \to 0} \int\limits_{-\delta k}^{\delta k}{\rm 
d}{k}\,b_{\epsilon}({k}). \]
Skipping some lengthy but elementary calculations, we produce 
only the final 
result:	 \begin{equation}
\dot\Phi=2(1-h\cos\Phi)-w,
\label{n13a}
\end{equation}
\begin{equation}
\dot{w}=-\Bigl\{(\gamma+h\sin{\Phi })
(4-\rho )-2{{\sigma}}  h\cos{\Phi }\Bigr\}w,
\label{n13b}
\end{equation}
\begin{equation}
{\dot \rho}=\left\{ \frac{w}{2}-{{\frak c}} H
+({{\frak c}}+2)h\cos{\Phi }\right\} {{\sigma}}
-(\gamma+h\sin{\Phi })(4-{\rho} ),
\label{n14a}\end{equation}
\begin{equation}
{\dot \sigma}=\left(-\frac{w}{2}+\frac{H}{3}
+\frac{1}{3}h\cos{\Phi }\right){\rho} +
(\gamma+3h\sin{\Phi }){{\sigma}} .
\label{n14b}\end{equation}
Here $w= 8 \eta^2$ and ${{\frak c}} =4\ln2-2\approx0.77$.

Below we will compare conclusions based on the analysis of the  
four-dimensional system (\ref{n13a})-(\ref{n14b}) with
solutions of the full nonreduced NLS equation.
To facilitate the comparison with results available in literature,
we take the damped-driven NLS in autonomous form:
\begin{equation}
i\psi_{t}+\psi_{xx}+2|\psi|^2\psi=(1-i\gamma)\psi+h {\psi}^*.
\label{standard}
\end{equation}
Here $\psi$ is related to a solution of  equation (\ref{equ1})
by a simple phase transformation:
$\psi(x,t)=-\exp(-i\Omega t) u (x,t)$.
The value of the field (which is a sum of the soliton
and radiation) at the centre of the soliton, i.e., at $x=0$,
can be expressed through the variables of the reduced system.
Using (\ref{solution}) and~(\ref{qr}) we obtain in the limit
$\epsilon\rightarrow0$:
\[
\left.  \phantom{\frac11} u (x,t)\right|_{x=0}=2i\eta \exp
\left(i\Omega t -\frac{i}{2}\Phi\right)
\left(1+\frac{i{{\sigma}} -{\rho} }{2\pi}
\right),
\]
and hence the value of the solution of the
NLS (\ref{standard}) at the point $x=0$ is given by
\begin{equation}
\left. \phantom{\frac11} \psi(x,t) \right|_{x=0}  =
\exp\left\{-i\left(\frac{\Phi +\pi}{2}\right)\right\}
\sqrt{\frac{w}{2}}
\left(1+\frac{i{{\sigma}} -{\rho} }{2\pi}\right)
\equiv \psi_0(t).
\label{n17}
\end{equation}

The finite-dimensional system has two exact solutions.
One of these,
\begin{equation}
\Phi =-\pi + \arcsin(\gamma/h), \quad w=2(1+H), \quad
{\rho} ={{\sigma}} =0,
\label{fixed_point}
\end{equation}
corresponds to the pure-soliton solution of  equation
(\ref{standard}). The second solution arises by letting  $w=0$, with 
$\Phi$ determined from the equation (\ref{n13a}), and  $\rho$ and $\sigma$
defined by the linear system (\ref{n14a}) - (\ref{n14b}).
Choosing the constant of  integration so that  $\Phi (0)=0$, we obtain
\begin{equation}
\Phi (t)=2\arctan\left\{\left(\frac{1-h}{1+h}\right)^{\frac{1}{2}}
\tan(\sqrt{1-h^2}\,t)\right\} \; {\rm mod\;} (2 \pi).
\label{n18}\end{equation}
Substituting this into equations (\ref{n14a}) and (\ref{n14b}), the  system 
for  ${\rho} $ and ${{\sigma}}$ becomes
\[
\dot\rho =(\gamma+h d_1)({\rho}-4)
+( 4 \ln2 \, h d_2 - {\frak c}) {{\sigma}},
\]
\begin{equation}
 \label{radzero}\end{equation}
\[
\dot\sigma=(\gamma+3h d_1){{\sigma}}
+\frac{1}{3}(H - h d_2){\rho},
\]
where
\[
d_1 = \sin\Phi(t)=\frac{\sqrt{1-h^2}\sin(2\sqrt{1-h^2}t)}
{1+h\cos(2\sqrt{1-h^2}t)},
\quad
d_2 = \cos\Phi(t)=\frac{h+\cos(2\sqrt{1-h^2}t)}
{1+h\cos(2\sqrt{1-h^2}t)}.
\]

\begin{figure}
\psfig{file= 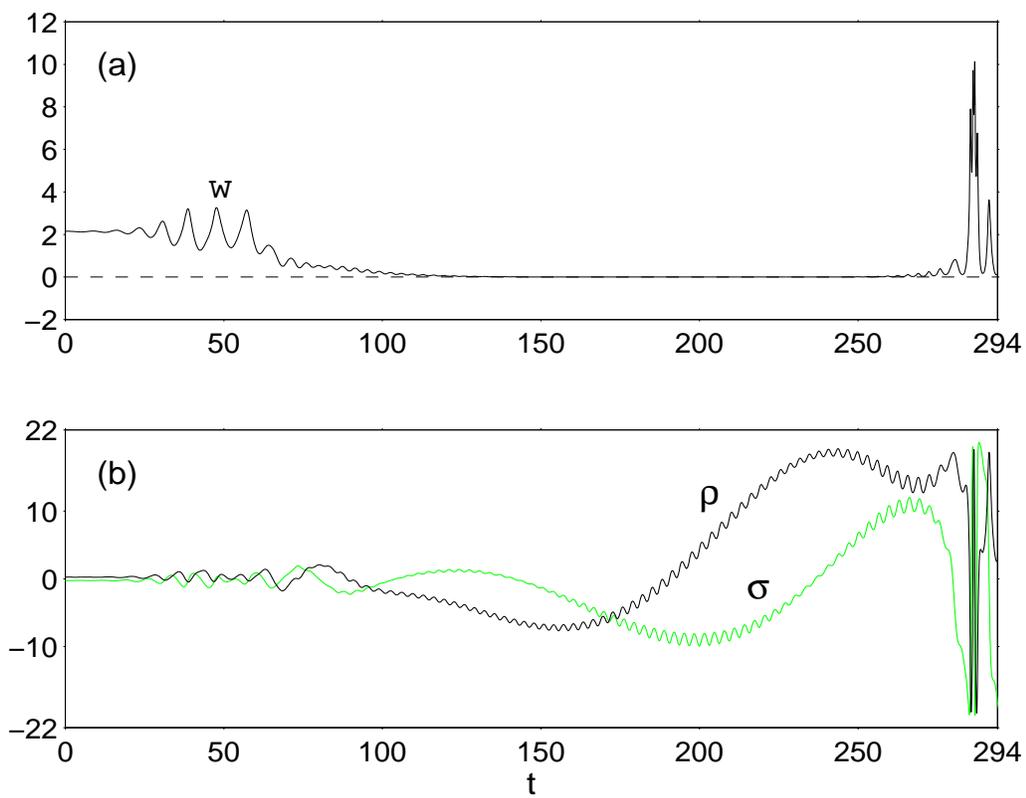,height=0.65\textwidth,width=0.85\textwidth}
\caption{\sf A numerical evolution of $w$, $\rho$, and $\sigma$:
an attraction to the
$w=0$ solution and an explosive instability of the latter due to the
exponential growth of $\rho$ and $\sigma$.}
\label{wrhosigm}
\end{figure}

In view of equations (\ref{vari_sol}) and (\ref{qr}), $w=0$ amounts to
$\psi(x,t)=u(x,t)=0$ and hence the above $w=0$ solution  corresponds
to the flat zero solution of the full nonreduced NLS equation
(\ref{standard}). This fact is quite remarkable. Indeed, although the 
system (\ref{n13a})-(\ref{n14b}) was derived under the assumption of the 
proximity to the (damped driven) soliton, it does possess the flat solution 
which cannot be  regarded  as the soliton's small perturbation. As in the 
full partial differential equation (\ref{standard}), where the unstable
soliton can decay to the flat attractor, finite-dimensional
trajectories starting near the unstable fixed point (\ref{fixed_point})  
can be attracted to the $w=0$ solution. An example of such evolution,
obtained numerically,  is presented in Fig.~\ref{wrhosigm}, for $t \in [0,250]$.

It is important to note here that for $\gamma>0$, solutions to the linear
system (\ref{radzero}) are exponentially growing functions
(see Fig.~\ref{wrhosigm}(b)). This is not in disagreement with the fact 
that the corresponding $\psi(x,t)$ is zero for all $x$ and $t$.
Indeed, the total amplitude of the long-wavelength radiation 
includes a factor 
$\eta$ (see e.g. (\ref{qr}) or (\ref{n17}).)
 Therefore, 
if  $w=\eta=0$, the total amplitude is zero no matter what ${\rho} $ and 
${{\sigma}} $ are equal to. 

The fact that $\rho$ and $\sigma$ are exponentially growing
functions, gives rise to an instability of the $w=0$ solution
of the system (\ref{n13a})-(\ref{n14b}). Indeed, the multiplier
\[
{\frak r}(t)= -\left\{
(\gamma + h \sin \Phi) (4 - \rho) - 2 \sigma h \cos \Phi \right\}
\]
in the right-hand side of (\ref{n13b}), may assume large positive values.
(For example, each time $\sin \Phi$ goes through 1, ${\frak r}(t)$
becomes equal to approximately $(\gamma+h) \rho$.)
This means that for $w \, ^< \!\! \, \!\!_\sim \, 1$ we have two
different time scales in the system: the variables $\Phi$,
$\rho$ and $ \sigma$ do not change appreciably on intervals
$\Delta t \, ^< \!\! \, \!\!_\sim \, 1$ and can be considered
constant, while $w$ grows with the exponential growth rate ${\frak r} \gg 1$.
Consequently, if $w$ is assigned a small but nonzero value
at the moment of time when ${\frak r}(t)$ is large, it will quickly
(within $\Delta t \sim {\frak r}^{-1}$) grow to values of order 
${\frak r}$. This is
indeed observed in simulations, see Fig.\ref{wrhosigm} for $t> 250$.
However, the instability of the $w=0$  solution  is  spurious ---
in the sense that it does not mirror any genuine instabilities of the  zero 
solution in the full  NLS equation.

\section{Reduced finite-dimensional dynamics}
\label{fourth}
\subsection{Onset of instability}

Linearising the system (\ref{n13a})-(\ref{n14b})  about the fixed point
(\ref{fixed_point}) and using (\ref{approxi}), we obtain
\begin{equation}
{\ddot \phi} +2\gamma {\dot \phi}
+8 H (1+ H ){\phi }=4 H (1+ H ){{\sigma}} ,
\label{s1a}\end{equation}
\begin{equation}
{\ddot \sigma}
+2\gamma {\dot \sigma}+(1+ H )[1-(2{\frak c}+1) H ]{\sigma} =
-4 H (1+ H ){\phi }.
\label{s1b}
\end{equation}
Here $\phi$ is the perturbation of the soliton's phase, defined as 
$\phi=\Phi-\Phi_+$. Letting ${\phi}(t)=\chi e^{\lambda t}$ and 
${{\sigma}}(t) ={{\sigma}}_0 e^{\lambda t}$, where $\lambda =  -\gamma+ i 
\omega$, we obtain the characteristic equation for  complex $\omega$: 
\begin{equation} {\omega}^4 -(\Omega_s^2+\Omega_r^2-2\gamma^2){\omega}^2
+(\Omega_s^2-\gamma^2)(\Omega_r^2-\gamma^2)+16H^2(1+ H )^2=0.
\label{s2}
\end{equation}
Here
\[
\Omega_s^2=8H(1+H), \quad
\Omega_r^2=(1+H)[1-(2{{\frak c}}+1)H].
\]

Like  the  eigenvalue problem (\ref{ch1})-(\ref{ch3}) for the full set of 
spectral data (and like the eigenvalue problem for the underlying partial 
differential equation \cite{M5}), the characteristic equation (\ref{s2})
can be conveniently reformulated in terms of  the self-similar variable 
$\mu = \gamma^2 + \omega^2$: 
\begin{equation} {\mu}^2 - p \mu +q =0,
\label{s3} \end{equation}
where
\[
p=p(H)= \Omega_s^2 +\Omega_r^2, \quad
q=q(H)=\Omega_s^2 \Omega_r^2 + 16H^2(1+ H )^2.
\]
The fact that the reduced system (\ref{n13a})-(\ref{n14b})
inherits the self-similarity of the parent PDE, deserves to be specially
emphasised. It implies that the reduction procedure based
on the Riemann-Hilbert problem --- unlike the variational
reductions \cite{NEEDS,Longhi_var} --- preserves the
structure of the infinite-dimensional phase space.

The fixed point is unstable when ${\rm Re\/} \lambda>0$. In terms of $\mu$, 
this condition translates into 
\begin{equation}
\frac{[{\rm Im\/} \mu(H)]^2}{4 {\rm Re\/} \mu(H)} \ge \gamma^2.
\label{onset}
\end{equation}
The discriminant of (\ref{s3}) is
\[
D(H)=(H_1 H_2)^{-1} (1+H)^2(H-H_1)(H-H_2),
\]
where $H_1= (8 \ln 2 +13)^{-1} \approx0.054$ and $H_2 = (8 \ln 2 -3)^{-1} 
\approx0.39$. Between  $H_1$ and $H_2$ the quadratic (\ref{s3}) has a pair
of complex-conjugate roots while outside this interval both roots
are real and nonnegative (Fig. 3). Consequently, the inequality 
(\ref{onset}) can only hold for $H_1 \le H \le H_2$. For $H$ within this 
interval the inequality (\ref{onset}) amounts to 
\begin{equation}
8 \gamma^2 \le -D(H)/p(H).
\label{eight_gamma}
\end{equation}
For small $\gamma$ the cubic equation $D(H)/p(H)=-8\gamma^2$
has one negative and two positive roots which we denote
${\cal H}_1(\gamma)$ and ${\cal H}_2(\gamma)$,
$0< {\cal H}_1(\gamma) < {\cal H}_2(\gamma)$.
(Note that ${\cal H}_1(0)=H_1$ and ${\cal H}_2(0)=H_2$; hence the
notation.) Recalling that $H=\sqrt{h^2-\gamma^2}$, the instability 
inequality (\ref{eight_gamma})  can be rewritten as
\begin{equation}
\sqrt{{\cal H}_1^2(\gamma)+ \gamma^2} \le h \le
\sqrt{{\cal H}_2^2(\gamma)+ \gamma^2}.
\label{H1H2}
\end{equation}
As $\gamma$ is increased, the positive roots ${\cal H}_1(\gamma)$
and ${\cal H}_2(\gamma)$ merge and become complex. For greater $\gamma$
the inequality (\ref{onset})-(\ref{eight_gamma}) cannot be satisfied
by any $h$.

Equation (\ref{H1H2}) gives an explicit form of the
instability region on the $(h,\gamma)$-plane (Fig.4). As
$h$ is increased past the value $h=\sqrt{{\cal H}_1^2+ \gamma^2}$
(or decreased below $h=\sqrt{{\cal H}_2^2 + \gamma^2}$), the fixed
point becomes unstable via a Hopf bifurcation.

\begin{figure}
\psfig{file= 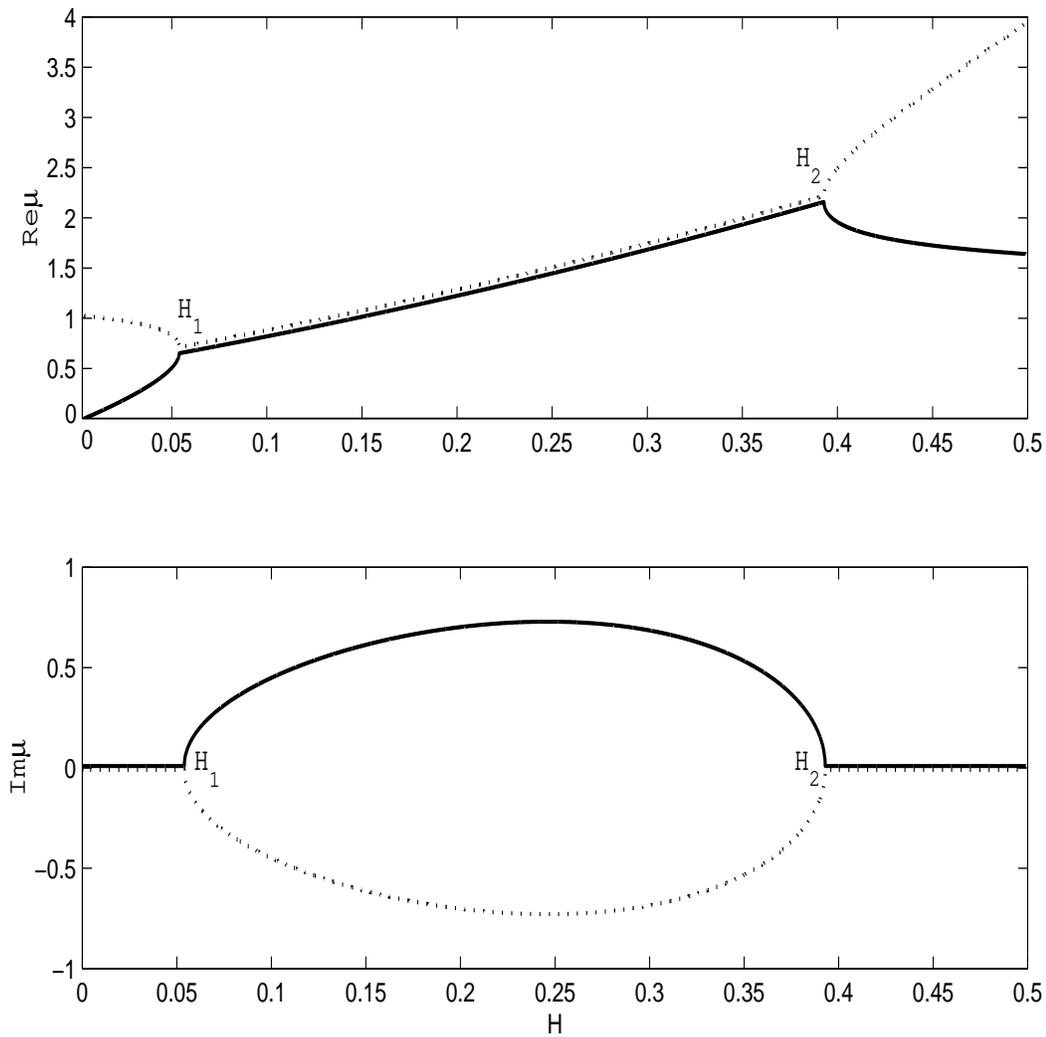,height=0.85\textwidth,width=0.85\textwidth}
\caption{\sf The real (a) and imaginary (b)
parts of the roots to the characteristic
equation (\ref{s3}).}
\end{figure}

\begin{figure}
\psfig{file= 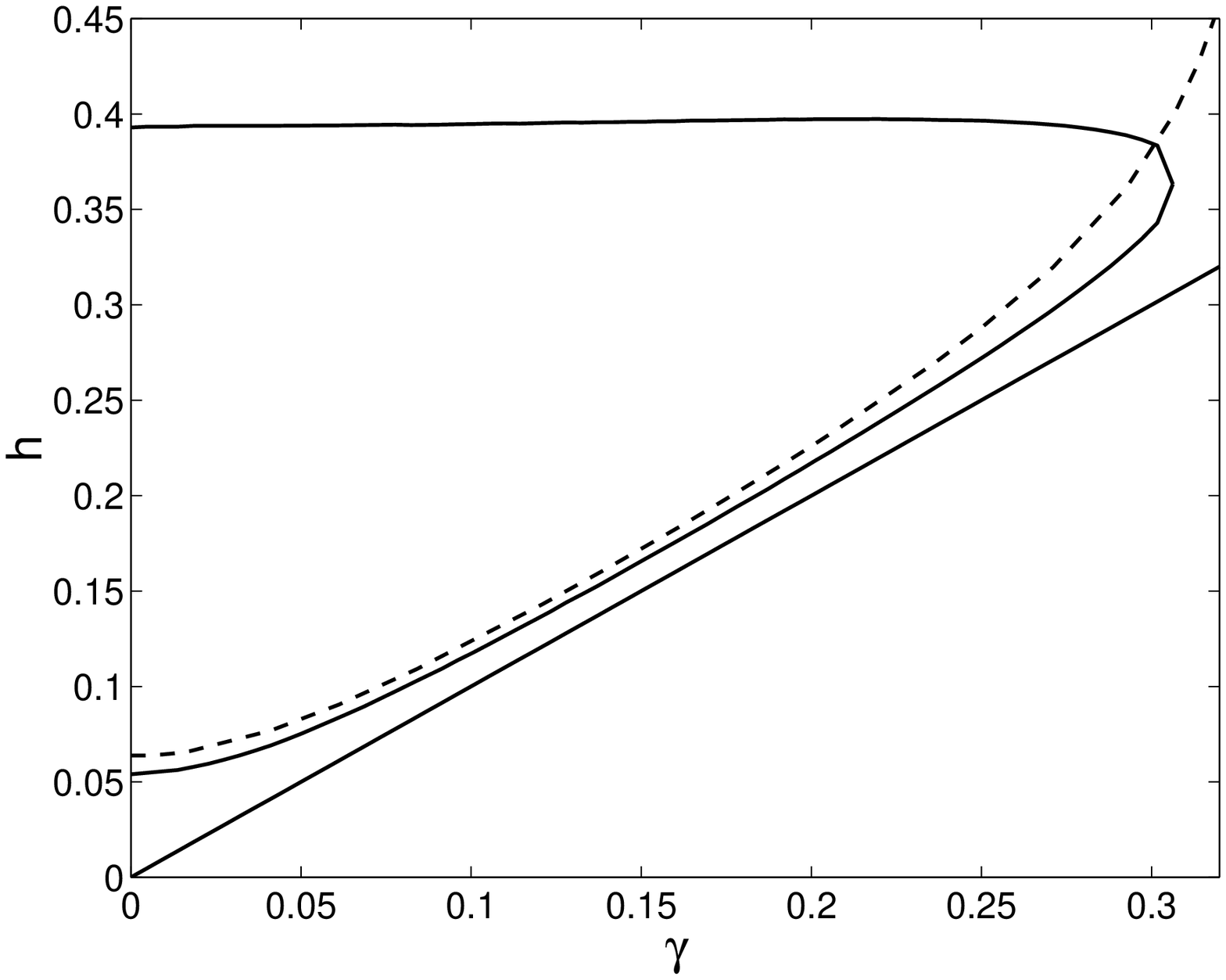,height=0.5\textwidth,width=0.85\textwidth}
\caption{\sf
Solid curve: the  boundary of the instability
region of the fixed point. (The fixed point is unstable {\it inside\/} the
``parabola".) Dashed curve: the line of the Hopf bifurcation of the
stationary soliton of the full NLS
equation (\ref{standard}). (The soliton
is unstable {\it above\/} the dashed curve.)
Also shown is the straight line $h= \gamma$, the boundary of the existence 
domain of both the soliton and the fixed point (\ref{fixed_point}).}
\end{figure}

The  instability setting in for  $H > {\cal H}_1(\gamma)$
corresponds to the Hopf instability of the soliton in the full
damped-driven
NLS equation.
Note however that the finite-dimensional  instability threshold 
$h=\sqrt{{\cal H}_1^2(\gamma)+ \gamma^2}$ is somewhat lower than the  
instability threshold in the partial differential equation (which is 
also shown in 
Fig. 4).  On the other hand, the upper boundary of the instability region,
$h=\sqrt{{\cal H}_2^2(\gamma)+\gamma^2}$, does not have a counterpart in
the full NLS equation. (In the full PDE, the soliton does not restabilise 
as $h$ is increased \cite{M5}.) This fact alone is sufficient to conclude 
that the finite-dimensional system {\it cannot\/} be expected to provide a 
good approximation to the infinite-dimensional dynamics for $h$
greater than approximately $0.4$. 

In order to find attractors in the region inside the parabola (\ref{H1H2}), 
where the fixed point is unstable, we performed a series of numerical 
simulations of equations (\ref{n13a})-(\ref{n14b}). Here our strategy was 
similar to that used in the simulations of the full nonreduced NLS equation
\cite{Bondila}; that is,  we varied $h$ for a fixed  value of $\gamma$.
We also adopted the same strategy with regard to the choice of the initial 
conditions. Our simulations always started with the (unstable) fixed point 
perturbed by values of order $10^{-14}$ which is several orders of
magnitude
 smaller
than the local discretisation error of the Runge-Kutta approximation.
(For some values of $h$ and $\gamma$ the final state of the system
was extremely sensitive to tiny changes of this perturbation.)

\subsection{Finite-dimensional attractors; $\gamma \neq 0$}
As $h$ is increased  for a fixed {\it nonzero\/}  $\gamma$
(with $\gamma \le 0.31$), a limit cycle is born supercritically at the 
point of the Hopf bifurcation, $h=\sqrt{{\cal H}_1^2(\gamma)+\gamma^2}$. 
This is in exact agreement with  the full partial differential equation.  
The subsequent bifurcation diagram depends on the value of $\gamma$.

\begin{center} 
\begin{figure}
\psfig{file= 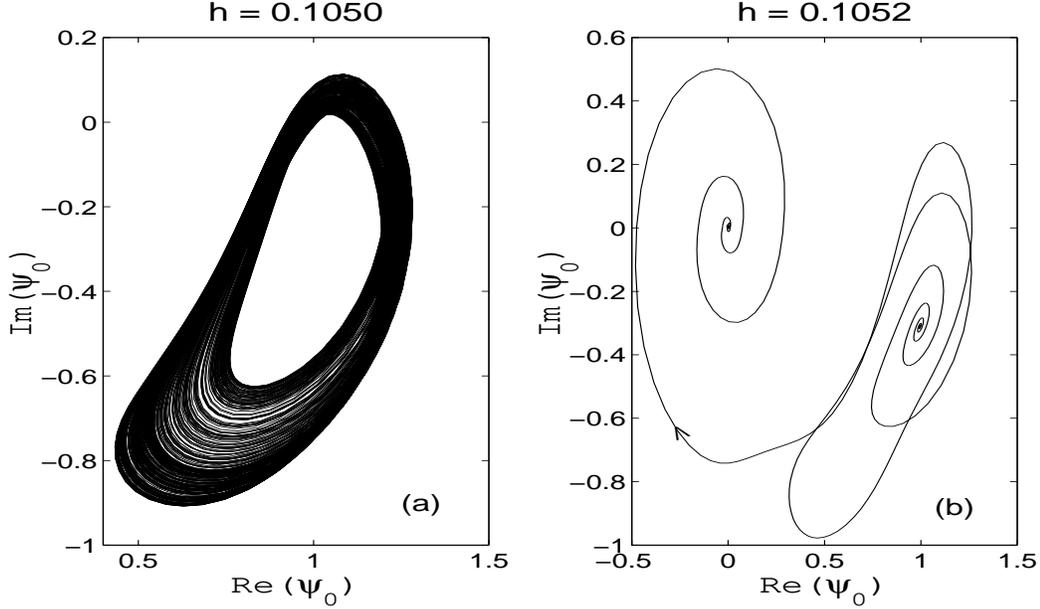,height=0.5\textwidth,width=0.85\textwidth}
\caption{\sf Trajectories in the four-dimensional system
(\ref{n13a})-(\ref{n14b}) with $\gamma=0.06$. (a)  A strange attractor
corresponding to the chaotic soliton. (b) A crisis of the chaotic 
attractor: the trajectory escapes from the neighbourhood of the fixed point 
and flows into the $w=0$ solution.  To facilitate the comparison with
the corresponding regimes in the PDE, the trajectories are presented
in terms of the real and imaginary parts of the quantity
(\ref{n17}).} 
\label{zero_attractor}
\end{figure}
\end{center}

For  $\gamma$ greater than approximately 0.2, the 
stable limit cycle persists as
$h$ is increased all the way up to $h=\sqrt{{\cal H}_2^2(\gamma)+\gamma^2}$ 
where it shrinks back to the  fixed point. On the contrary, if we increase 
$h$  for a smaller $\gamma$, $0.12 < \gamma \le 0.20$, the limit cycle 
looses its stability at a certain value  $h=h_{\rm z}$
(where $h_{\rm z} < \sqrt{{\cal H}_2^2+ \gamma^2}$.)
 For $h$ above $h_{\rm z}$, 
the finite-dimensional trajectory emanating from {\it any\/} initial 
condition, quickly settles to the $w=0$ solution corresponding to the zero 
attractor of the full nonreduced NLS. (This is illustrated by 
Fig.\ref{wrhosigm} and, for a different $\gamma$,
Fig.\ref{zero_attractor}(b).) Increasing $h$ still further, the $w=0$ 
solution persists as the only attractor over a sizeable range of $h$ values 
--- until a stable limit cycle reappears. This range becomes wider for 
smaller values of $\gamma$.

The range of driving strengths where the only attractor is
$w=0$, exists for {\it all\/} $\gamma \in (0, 0.20]$
(although for some $\gamma$ the limit cycle may undergo a number of
intermediate bifurcations before disappearing from the attractor chart;
see below.) This is in exact agreement with the behaviour observed
in the full partial differential equation \cite{Bondila} where the
solitonic attractor undergoes a crisis and the flat zero solution remains
the only attractor in the system. Thus we may conclude that taking into
account the coupling of the soliton to the $k=0$ radiation,
is sufficient to explain the occurrence of the ``desert region"
on the $(h, \gamma)$-plane. It is appropriate to mention here that
a similar ``desert" spanned just by a flat attractor, arises in the
{\it externally\/} driven NLS \cite{add1,SPLE}. It is natural
to assume that the appearance of the ``flat desert" in the
latter system can be still explained by the soliton-longwave
radiation coupling. (The failure of the four-dimensional reduced systems
proposed in \cite{add1,SPLE,TSFGR} to capture the crisis of the
localised attractor and reproduce the ``desert", should be
probably attributed to nonoptimal variational Ans\"atze.)

Although our finite-dimensional system with $ \gamma \le 0.2$
correctly reproduces the sequence of attractors arising in  the full PDE 
(stationary soliton $\to$ oscillating soliton $\to$  zero solution), it does
not necessarily capture  fine details  of this sequence. Unlike
the oscillating soliton in the full NLS equation, for $\gamma$ between 0.12 
and 0.20 the finite-dimensional limit cycle does not undergo any 
period-doubling bifurcations. The largest $\gamma$ for which the 
period-doubling occurs in the reduced system (\ref{n13a})-(\ref{n14b}), is 
$\gamma=0.12$. In this case the period-2 cycle arises at $h= 0.16$
and then degenerates back to the period-1 as $h$ is increased past
$h=0.17$. Increasing $h$ still further, the period-1 yields to the $w=0$
solution for $h \ge 0.195$, without any intermediate period-doublings.
A similar pattern arises for $\gamma=0.11$.

In the case $\gamma=0.10$ the sequence of finite-dimensional attractors is 
richer. In this case we observed the whole cascade of period-doubling 
bifurcations,  culminating, for $h=0.146$, in a chaotic attractor (centred 
on the fixed point). For $h$ between 0.146 and 0.150 the only attractor was
found to be $w=0$; for $h=0.150$ the chaotic attractor reappears,
and for even greater $h$ it degenerates to the period-2  (for $h=0.165$) 
and then period-1 cycle (for $h=0.170$). At $h = 0.18$ the cycle disappears 
and $w=0$ was the only attractor we detected in a wide band of $h$ values 
above $h=0.18$. Next, increasing $h$ for the fixed  $\gamma=0.05$ and 
0.07, the period-1 ($P1$) limit cycle yields to the period-2 ($P2$) and 
then to $w=0$. For $\gamma=0.06$ the sequence of attractors was $P1 \to$ 
$P2 \to$ $(w=0) \to $ $P3 \to$ $P4 \to$  (a 4-band chaotic attractor) $\to$ 
$(w=0) \to $ (a 2-band chaos coexisting with $P3) \to$  (a 1-band chaos) 
$\to$   $(w=0)$. (The last two regimes are illustrated in  
Fig.\ref{zero_attractor}.) The full NLS equation also exhibits a rich host 
of attractors for $\gamma > 0.05$, including the chaotic soliton 
\cite{Bondila}, \cite{Bondila2}; however, details of the two bifurcation 
diagrams do not necessarily coincide.

It is only for very small $\gamma$ that the finite- and 
infinite-dimensional dynamics are in exact agreement. Increasing $h$
for the fixed $\gamma=0.01, 0.02$ and 0.04, the period-1 cycle yields 
directly to the $w=0$ attractor. This is consistent with the absence of the 
period-doubling in the full PDE. The smallest $\gamma$ for which
the period-2 soliton was observed there, was $\gamma=0.055$ \cite{Bondila2}.

\subsection{The undamped case, $\gamma=0$}
\begin{center}
\begin{figure}
\psfig{file= 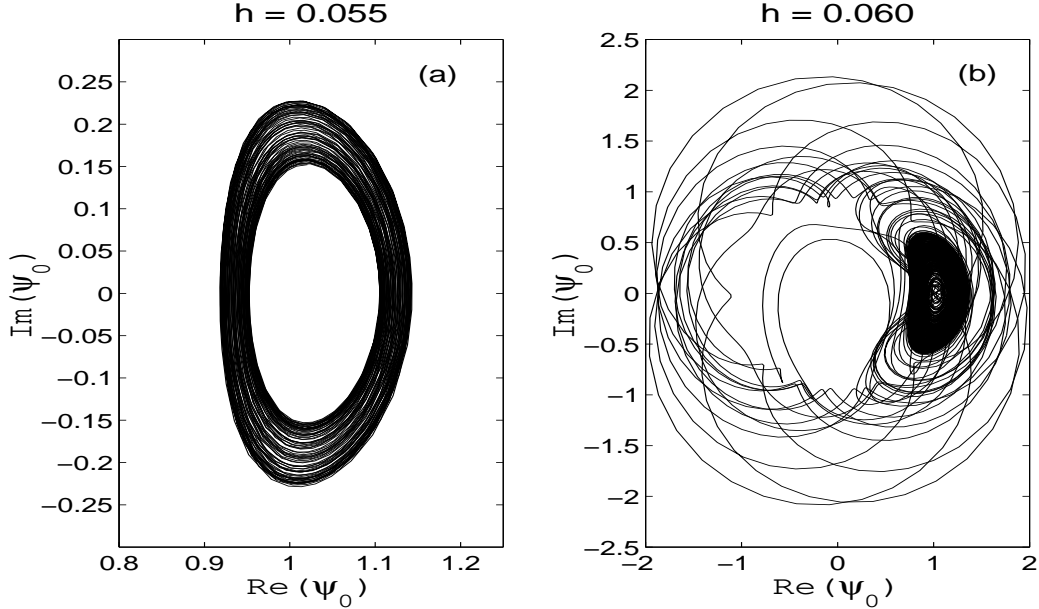,height=0.5\textwidth,width=0.85\textwidth}
\caption{\sf Trajectories in the finite-dimensional system
(\ref{n13a})-(\ref{n14b}) with $\gamma=0$.   The strange attractor
(a) and its crisis (b).}
\label{chaotic_solution}
\end{figure}
\end{center}

This case deserves a separate consideration as the
arising attractors are very different from those occurring
for nonzero damping. Consistently with the PDE \cite{add2}, no limit cycles 
were detected for $\gamma=0$. Instead, we observed two types of chaotic 
regimes. For $h$ in a narrow interval $ 0.0539< h< 0.06$ just above the 
Hopf bifurcation value $h=H_1=0.0539$, the trajectory was seen to wind 
chaotically in an annulus centred on  the unstable fixed point 
(Fig.\ref{chaotic_solution}a). The outer radius of the annulus is 
approximately one order of magnitude smaller than $w=2(1+h)$, the 
$w$-coordinate of the fixed point. This proximity of the orbit to the fixed 
point justifies one of the assumptions made in the derivation of the 
finite-dimensional system (\ref{n13a})-(\ref{n14b}). The  strange attractor 
shown in Fig.\ref{chaotic_solution}a represents small chaotic oscillations
of the amplitude of the  soliton about its stationary value  $\left| \psi 
\right|_{x=0}=\sqrt{1+h}$. The chaotic oscillations of the undamped soliton 
were indeed observed in numerical simulations of the full nonreduced NLS 
equation \cite{Nora}; however these would die out as  transients and the 
evolution settle to either a slowly growing or decaying breather \cite{add2}.
The fact that the chaotic attractor does not arise in the full
partial differential equation with $\gamma=0$ and only persists as a 
transient chaos, can be explained by the interaction of the soliton with
short and medium wavelength radiations --- which we have neglected
in our present derivation of the finite-dimensional system. 
For $h$ greater than 0.06 (but smaller than $H_2=0.39$, the upper
boundary of the fixed point's instability domain), the chaotic solution
is no longer  centred on the fixed point. (See Fig.\ref{chaotic_solution}b).
What is even more important, the amplitude of the aperiodic oscillations
grows rapidly. Therefore the finite-dimensional system is not applicable
in this region and the chaotic solution shown in Fig.\ref{chaotic_solution}b
does not have a counterpart in the full nonreduced partial differential 
equation.

\section{Conclusions}
\label{fifth}
It was proposed by various authors \cite{add2,add1,TSFGR,Birnir,Bishop}
that the coupling to the radiation (more specifically, to its 
long-wavelength component \cite{add1,TSFGR,Birnir,SPLE})
is the key factor determining the internal dynamics of the damped driven
soliton, in particular its instability, bifurcations and transition to
chaos. The main objective of the present work was to study
the role of the soliton-radiation interaction within an approach which
allows a rigorous decomposition of the phase space into the soliton and
radiation modes. Our treatment is based on the Riemann-Hilbert problem,
a modern version of the Inverse Scattering transform.  Although the
Inverse Scattering method had already been utilised in a related context
(for the {\it externally\/} driven NLS \cite{9,add1,TSFGR,Bishop}), our
approach is different in that we are {\it not\/} assuming  the damping and 
driving terms to be small in any sense. Instead, we are exploiting a 
remarkable coincidence between the mathematical formula for the 
parametrically driven damped soliton and that of the soliton  of the 
unperturbed NLS. This coincidence allows to associate the stationary 
damped-driven soliton with a stationary zero of the Riemann-Hilbert problem 
--- for {\it any\/} $h$ and $\gamma$. The evolution of all nearby
 solutions $\psi(x,t)$ can therefore be studied through  the evolution of 
the corresponding scattering data.

Some interesting insights are gained already from the linearised
equations for the spectral data. Conclusions of this analysis are
consistent  with results of the linearisation in the space of fields
$\psi(x,t)$ \cite{M5}; however the linearisation in the spectral space has 
an important advantage that it can be carried out analytically whereas
linearised equations for $\psi(x,t)$ could only be studied by means
of computer. The stationary soliton looses its stability as a result of its 
coupling to radiation waves. This had already been proposed before 
\cite{NEEDS} but now we have put this claim on rigorous footing. 

We attempted to advance beyond the linear approximation and track the
effect of the long-wavelength radiation on the {\it nonlinear\/}
dynamics of the soliton. To find a closed system for the evolution of
the scattering data, we had to make two assumptions. First, we assumed
that the solution of our damped driven NLS remains close to the
stationary soliton of the same, perturbed, equation.  Second, we 
assumed that the
resulting system for the evolution of the spectral data is linear in
radiation. (Note that we are {\it not\/} requiring the perturbation
in the right-hand side of (\ref{1}) to be small in any sense.)

The outcome of this analysis is a dynamical system
(\ref{n13a})-(\ref{n14b})
comprising  equations for the amplitude and
phase of the soliton, and for the complex amplitude of the $k=0$-radiation.  
Finite-dimensional   reductions of both externally 
\cite{add1,TSFGR,Birnir,SPLE} and parametrically \cite{NEEDS,Longhi_var} 
driven NLS are available in literature and hence we need to emphasise the 
differences. The principal difference between our  reduction and those 
obtained variationally \cite{TSFGR,SPLE,NEEDS,Longhi_var} or by the Galerkin 
projections \cite{Birnir}, lies in that our system of ODEs is not a product 
of any phenomenological Ansatz. It results from a rigorous expansion of 
the solution of the PDE over a set of ``nonlinear modes" and then retaining 
only those modes whose effect on the dynamics we are trying to track down. 
A failure of one or the other variational or Galerkin approximation to 
capture essentials of the supercritical dynamics of the soliton does not 
provide any information on {\it why\/} this particular Ansatz fails. One 
has to try a variety of different Ans\"atze, select the one that gives the 
best fit and then attempt to make some semi-intuitive conclusions on the 
role of this or that ingredient of the trial function. On the contrary, our 
approach allows to explore, systematically, each part of the phase space 
and identify the nonlinear modes responsible for each particular dynamical 
effect. Next, our reduction technique is different from the approach of an 
influential paper \cite{add1} which is also partially based on the Inverse 
Scattering. Besides the fact that the analysis of Ref.~\cite{add1}  relies 
on the smallness of the perturbation, it only uses the Inverse Scattering 
transform to obtain the functional form of the radiation wave whereas its
interaction with the soliton is introduced variationally.  The damping term 
for the radiation was not part of the variational algorithm and had to be 
added in an {\it ad-hoc\/} way. The method does not define the amplitude of 
the radiation either; this is introduced empirically
  and then fitted to match the
numerical data. Unlike Ref.~\cite{add1}, our finite-dimensional reduction 
is uniquely defined by the choice of the ingredients of the localised 
attractor and does not require introduction of  any phenomenological terms 
or fitting parameters. This uniqueness is reflected by the fact that (the 
linearisation of) our reduced system retains the self-similarity invariance 
of the (linearised) PDE.

The analysis of the reduced dynamical system shows that it is capable
of explaining only some parts and only some rough features of the
attractor chart of the parametrically driven damped NLS \cite{Bondila}.
Most notably, the interaction with long radiation waves is sufficient to
reproduce the approximate sequence of attractors arising when the driving
strength is increased under the fixed dissipation coefficient.
Consistently with computer simulations of \cite{Bondila,Bondila2}, the
finite-dimensional system exhibits the sequence
``stationary soliton $\to$ periodically
 oscillating soliton"
for larger dampings and ``stationary soliton $\to$ oscillating soliton
 $\to $ flat zero solution" for smaller $\gamma$. For $\gamma=0$
 the reduced system does not predict oscillating solitons with bounded
 amplitudes (apart from  a tiny window of $h$ values).
 This is also in agreement with the behaviour observed in the full
 PDE \cite{add2}. However, the finite-dimensional system
predicts --- erroneously
 --- the occurrence of the second, restabilising, Hopf bifurcation.
 As a result of that, the finite-dimensional fixed point turns out to be
 stable
 for all $\gamma$ as long as $ h \ge 0.4$, whereas the actual
 stability domain of the stationary soliton is much smaller \cite{M5}.
 One could have expected that the reduced and infinite-dimensional
 dynamics would be close in the region adjacent to the first,
 destabilising, Hopf bifurcation curve --- which does provide a reasonably
 good approximation to the Hopf bifurcation curve in the full PDE. In
 the actual fact, however, details 
 of the  attractors and bifurcation sequences
 in the
 two systems are quite different even in that region.
(An exception is the band of very small $\gamma$, $\gamma< 0.05$.)

Thus,
taking into account just the $k=0$
 component of radiation is insufficient for reproducing the entire
complexity
 in the damped-driven soliton's dynamics.
It is quite possible that the radiation waves with the frequency
close to the double frequency of the soliton's linear oscillations
\cite{add2} play a more important
(or equally important) role than those with $k=0$.
On the other hand, numerical simulations on finite intervals
reveal the excitation of radiations with {\it several\/}
 wavenumbers \cite{Bishop}. It is not unprobable
that a similar wavenumber selection and competition occur
on the infinite line. Finally, there are indications that the
oscillating soliton of the {\it externally\/} driven NLS,
is in fact a bound state of two solitons of the unperturbed, integrable,
NLS equation \cite{SPLE}. A similar mechanism may operate in
the parametrically driven case as well. Our approach allows to test
all these possibilities; we are planning to do so in future
publications.

\section{Acknowledgements}
We are grateful to Nora Alexeeva for her help in the course of this work.  
This project was supported by  grants from the Research Council of the 
University of Cape Town  and  the National Research Foundation of South 
Africa. 

\section{Appendix: Discrete eigenvalues of the operator
${\hat {\cal G}}$}
Here we demonstrate the existence of, and derive a lower bound  for, 
discrete eigenvalues of the operator ${\hat {\cal G}}$, equation 
(\ref{opG}). The smallest eigenvalue, $E_1$, can be sought for as a minimum 
of the corresponding Rayleigh quotient:
\begin{equation}
E_1=  \min_{\psi(k)} \frac{
\int {\rm d}{k} \psi {\hat {\cal G}} \psi}
{\int {\rm d}{k} \psi^2}.
\label{Rayleigh}
\end{equation}
In terms of $\varphi=(1+{k}^2) \psi$, this can be rewritten as
\begin{equation}
E_1=  \min_{\varphi(k)} \frac{
\int {\rm d}{k} \varphi
\left\{
(1+{k}^2)^{-2} \left[ (1+ \Omega {k}^2)^2-H^2 \right]
+ 2H \Omega {\hat {\cal F}} \right\}
\varphi}
{\int {\rm d}{k} (1+{k}^2)^{-2} \varphi^2},
\label{om_0}
\end{equation}
where
we have introduced a  symmetric operator
${\hat {\cal F}}$:
\begin{equation}
{\hat {\cal F}} \varphi({k}) = (1+{k}^2)^{-1} {\hat {\cal B}}
(1+{k}^2)^{-1} \varphi({k}).
\label{F_hat}
\end{equation}
(Here ${\hat {\cal B}}$ is given by (\ref{opR}).)

For $H=0$ the operator ${\hat {\cal G}}$ does not have
discrete eigenvalues but as $H$ grows,   at least one discrete
eigenvalue detaches from the
continuum.
Indeed,
in the region $|\ell|<1$, $|{k}|<1$ the kernel of
the operator  ${\hat {\cal F}}$,
\[
{\cal F}(\ell, {k})=
\frac{({{k}}-{\ell})({\ell}{{k}}-1)}
{\sinh\left[\frac{\pi}{2}({{k}}-{\ell})\right]}
\frac{1 }{ (1+{{k}}^2)^{3/2}(1+{\ell}^2)^{3/2} },
\]
satisfies
\begin{equation}
{\cal F}(\ell,k) \le \frac{1}{4 \sinh \pi} (\ell k-1).
\label{F_cal}
\end{equation}
 For an  arbitrary even function $h(k)$
the inequality (\ref{F_cal}) yields
\begin{equation}
\int\limits_{-1}^1\int\limits_{-1}^1{\rm d}k{\rm d}\ell\, h(k){\cal
F}(k,\ell)h(\ell)
< -\frac{1}{4\sinh{\pi}}
\left( \, \int\limits_{-1}^1{\rm d}kh(k)\right)^2.
\label{bound}
\end{equation}
Choosing a suitable trial function $\psi=\psi_H(k)$ in the quotient
(\ref{Rayleigh}) and using the estimate (\ref{bound}),
one readily concludes that for
any positive $H$ the quotient can take  values lying below
$1-H^2$, the edge of
the continuous spectrum. Hence the operator ${\hat {\cal G}}$ has at least
one discrete eigenvalue for $H>0$.

Next, the norm of the operator (\ref{F_hat}) is bounded:
\[
||{\hat {\cal F}} ||^2 =
\int\limits_{-\infty}^{\infty}
\int\limits_{-\infty}^{\infty}
{{\rm d}}{{k}}
{{\rm d}}{\ell} \, {\cal F}^2(\ell,{k})
< \frac58.
\]
Therefore ${\hat {\cal F}}$ is a Fredholm operator
and its eigenvalues ${\cal E}_m$, ${\hat {\cal F}} \phi_m=
 {\cal E}_m \phi_m$, are bounded:
\begin{equation}
{\cal E}_m^2 < ||{\hat {\cal F}} ||^2 < \frac58.
\label{58}
\end{equation}
Returning to the operator ${\hat {\cal G}}$ (\ref{opG}), its
 eigenvalue $E_n$ $(n=1,2,...)$  satisfies
\begin{equation}
E_n=   \frac{
\int {\rm d}{k} \psi_n {\hat {\cal G}} \psi_n}
{\int {\rm d}{k} \psi_n^2},
\label{ev_ef}
\end{equation}
where $\psi_n(k)$ is the associated eigenfunction:
${\hat {\cal G}}\psi_n=E_n \psi_n$. Using inequality (\ref{58})
in equation (\ref{ev_ef}) we obtain  a lower bound on $E_n$:
\begin{equation}
E_n >
 \frac{
\int {\rm d}{k} \varphi_n
\left(
 1-H^2
+ 2H \Omega {\hat {\cal F}} \right)
\varphi_n}
{\int {\rm d}{k}  \varphi_n^2}
>
(1-H^2)
\left[ 1-\frac{H\Omega}{1-H^2} \sqrt{\frac52} \, \right],
\label{om_1}
\end{equation}
where $\varphi_n=(1+k^2)\psi_n$.

\end{document}